\documentclass[twocolumn,trackchanges]{config/aastex701}
\usepackage{xcolor}

\usepackage{comment}
\usepackage{multirow}
\usepackage{subcaption}

\usepackage{amsmath}	
\usepackage{textgreek}
\usepackage{xargs}
\usepackage{xspace}

\let\oldAA\AA
\renewcommand{\AA}{\text{\oldAA}\xspace}
\newcommand{\mum}{\text{\textmu m}\xspace}

\newcommand{\Lyalpha}{\text{Ly\,\textalpha}\xspace}
\newcommand{\Halpha}{\text{H\,\textalpha}\xspace}
\newcommand{\Hbeta}{\text{H\,\textbeta}\xspace}
\newcommand{\Hgamma}{\text{H\,\textgamma}\xspace}
\newcommand{\Hdelta}{\text{H\,\textdelta}\xspace}

\newcommand{\Hepsilon}{\text{H\,\textepsilon}\xspace}

\newcommandx{\permittedEL}[6][1=O,2=III,3=,4=,5=,6=]{\text{{#1}\,{\sc {#2}}{#3}{#4}{#5}{#6}}\xspace}
\newcommandx{\semiforbiddenEL}[6][1=O,2=III,3=,4=,5=,6=]{\text{{#1}\,{\sc{#2}}]{#3}{#4}{#5}{#6}}\xspace}
\newcommandx{\forbiddenEL}[6][1=O,2=III,3=,4=,5=,6=]{\text{[{#1}\,{\sc{#2}}]{#3}{#4}{#5}{#6}}\xspace}

\newcommand{\HII}{\permittedEL[H][ii]}

\newcommandx{\NVL}[1][1=1243]{\permittedEL[N][v][\textlambda][#1]}
\newcommandx{\NVall}{\permittedEL[N][v][\textlambda][\textlambda][1239,][1243]}

\newcommandx{\CIIall}{\semiforbiddenEL[C][ii][\textlambda][\textlambda][2324--][2329]}

\newcommandx{\NIVL}[1][1=1486]{\semiforbiddenEL[N][iv][\textlambda][#1]}

\newcommandx{\CIVL}[1][1=1550]{\permittedEL[C][iv][\textlambda][#1]}

\newcommand{\HeII}{\permittedEL[He][ii]}
\newcommandx{\HeIIL}[1][1=1640]{\permittedEL[He][ii][\textlambda][#1]}

\newcommandx{\semiOIIIL}[1][1=1666]{\semiforbiddenEL[O][iii][\textlambda][#1]}

\newcommandx{\NIIIL}[1][1=1750]{\semiforbiddenEL[N][iii][\textlambda][#1]}

\newcommandx{\CIII}{\semiforbiddenEL[C][iii]}
\newcommandx{\CIIIL}[1][1=1909]{\semiforbiddenEL[C][iii][\textlambda][#1]}

\newcommandx{\NeIVL}[1][1=2424]{\forbiddenEL[Ne][iv][\textlambda][#1]}

\newcommandx{\MgIIL}[1][1=2803]{\permittedEL[Mg][ii][\textlambda][#1]}

\newcommandx{\NeVL}[1][1=3426]{\forbiddenEL[Ne][v][\textlambda][#1]}

\newcommand{\OIIperm}{\permittedEL[O][ii]}
\newcommand{\OII}{\forbiddenEL[O][ii]}
\newcommandx{\OIIL}[1][1=3726]{\forbiddenEL[O][ii][\textlambda][#1]}
\newcommandx{\OIIall}[2][1={3726,},2=3729]{\forbiddenEL[O][ii][\textlambda][\textlambda][#1][#2]}

\newcommandx{\NeIIIL}[1][1=3869]{\forbiddenEL[Ne][iii][\textlambda][#1]}

\newcommand{\OIII}{\forbiddenEL[O][iii]}
\newcommandx{\OIIIL}[1][1=5007]{\forbiddenEL[O][iii][\textlambda][#1]}
\newcommand{\OIIIall}{\forbiddenEL[O][iii][\textlambda][\textlambda][4959,][5007]}

\newcommandx{\NIL}[1][1=5200]{\forbiddenEL[N][i][\textlambda][#1]}

\newcommand{\OI}{\forbiddenEL[O][i]}
\newcommandx{\OIL}[1][1=6300]{\forbiddenEL[O][i][\textlambda][#1]}

\newcommandx{\HeIL}[1][1=5876]{\permittedEL[He][i][\textlambda][#1]}

\newcommand{\OIperm}{\permittedEL[O][i]}

\newcommandx{\OIresL}[1][1=8446]{\permittedEL[O][i][\textlambda][#1]}

\newcommand{\NII}{\forbiddenEL[N][ii]}
\newcommandx{\NIIL}[1][1=6583]{\forbiddenEL[N][ii][\textlambda][#1]}
\newcommand{\NIIall}{\forbiddenEL[N][ii][\textlambda][\textlambda][6548,][6583]}

\newcommand{\SII}{\forbiddenEL[S][ii]}
\newcommandx{\SIIL}[1][1=6716]{\forbiddenEL[S][ii][\textlambda][#1]}
\newcommandx{\SIIall}[2][1=6716,2=6731]{\forbiddenEL[S][ii][\textlambda][\textlambda][{#1},][#2]}

\newcommandx{\OIIAuL}[1][1=7325]{\forbiddenEL[O][ii][\textlambda][#1]}
\newcommand{\OIIAuall}{\forbiddenEL[O][ii][\textlambda][\textlambda][7319--][7331]}

\newcommandx{\CIIFIRL}{\forbiddenEL[C][ii][\textlambda][158\,\mum]}

\newcommand{\NaI}{\permittedEL[Na][i]}
\newcommandx{\NaIL}[1][1=5890]{\permittedEL[Na][i][\textlambda][#1]}
\newcommand{\NaIall}{\permittedEL[Na][i][\textlambda][\textlambda][5890,][5896]}

\newcommand{\CaII}{\permittedEL[Ca][ii]}
\newcommandx{\CaIIL}[1][1=3934]{\permittedEL[Ca][ii][\textlambda][#1]}

\newcommandx{\MgIb}{\permittedEL[Mg][i][b]}
\newcommandx{\FeIperm}{\permittedEL[Fe][i]}
\newcommandx{\FeIIperm}{\permittedEL[Fe][ii]}
\newcommandx{\TiIIperm}{\permittedEL[Ti][ii]}
\newcommandx{\FeIIpermL}[1][1=2600]{\permittedEL[Fe][ii][\textlambda][#1]}
\newcommandx{\FeII}{\forbiddenEL[Fe][ii]}
\newcommandx{\FeIIL}[1][1=5159]{\forbiddenEL[Fe][ii][\textlambda][#1]}
\newcommandx{\FeIIall}{\forbiddenEL[Fe][ii][\textlambda][\textlambda][4359,][4414]}
\newcommandx{\FeVL}[1][1=4071]{\forbiddenEL[Fe][v][\textlambda][#1]}
\newcommandx{\FeVIIL}[1][1=5160]{\forbiddenEL[Fe][vii][\textlambda][#1]}

\usepackage{etoolbox}
\makeatletter
\newcommand\sendemail[4]{
\edef\@tempa{mailto:#1?subject=#2&body=#3 }%
\edef\@tempb{\expandafter\html@spaces\@tempa\@empty}%
\href{\@tempb}{#4}}

\catcode\%=11
\def\html@spaces#1 #2{#1
\catcode\%=14
\makeatother

\newcommand{\todo}[1]{\textcolor{magenta}{[#1]}}
\newcommand{\orcid}[2]{\href{http://orcid.org/#2}{#1}}
\newcommand{\orcidsymb}[2]{\href{http://orcid.org/#2}{#1\adjustbox{trim={-.15\width} {0\height} {-.15\width} {0\height},clip}{\includegraphics[height=10pt]{figs/orcid.pdf}}}}

\newcommand{\citationneeded}{\textcolor{ForestGreen}{$^{\rm citation\;needed}$}}
\let\oldtextsigma\textsigma
\renewcommand{\textsigma}{\oldtextsigma\xspace}
\let\oldtextdegree\textdegree
\renewcommand{\textdegree}{\oldtextdegree\xspace}

\newcommand{\kms}{\ensuremath{\mathrm{km\,s^{-1}}}\xspace}
\newcommand{\Msun}{\ensuremath{{\rm M}_\odot}\xspace}
\newcommand{\Zsun}{\ensuremath{{\rm Z}_\odot}\xspace}
\newcommand{\yr}{\ensuremath{{\rm yr}}\xspace}
\newcommand{\Myr}{\ensuremath{{\rm Myr}}\xspace}
\newcommand{\Gyr}{\ensuremath{{\rm Gyr}}\xspace}
\newcommand{\peryr}{\ensuremath{{\rm yr^{-1}}}\xspace}
\newcommand{\Lsun}{\hbox{\,${\rm L}_\odot$}}
\newcommand{\kpc}{\text{kpc}\xspace}
\newcommand{\ZH}{\text{[Z/H]}\xspace}
\newcommandx{\percm}[1][1=3]{\ensuremath{\mathrm{cm}^{-#1}}\xspace}	

\newcommandx{\lambdar}[2][1=R,2=]{\ensuremath{\lambda_{\rm {#1}}{#2}}\xspace}
\newcommand{\eps}{\ensuremath{\epsilon}\xspace}
\newcommand{\mstar}{\ensuremath{M_\star}\xspace}
\newcommand{\mdyn}{\ensuremath{M_\mathrm{dyn}}\xspace}
\newcommand{\re}{\ensuremath{R_\mathrm{e}}\xspace}
\newcommand{\vstar}{\ensuremath{v_\star}\xspace}
\newcommand{\vnai}{\ensuremath{v_{\NaI}}\xspace}
\newcommand{\sigmastar}{\ensuremath{\sigma_\star}\xspace}
\newcommand{\sigmaestar}{\ensuremath{\sigma_{\star,\mathrm{e}}}\xspace}
\newcommand{\vperc}[1]{\ensuremath{v_{#1}}\xspace}

\newcommand{\vesc}{\ensuremath{v_\mathrm{esc}}\xspace}
\newcommand{\nelec}{\ensuremath{n_\mathrm{e}}\xspace}
\newcommand{\Telec}{\ensuremath{T_\mathrm{e}}\xspace}
\newcommand{\Rout}{\ensuremath{R_\mathrm{out}}\xspace}
\newcommand{\vout}{\ensuremath{v_\mathrm{out}}\xspace}
\newcommandx{\Mout}[2][1=,2=]{\ensuremath{M_{\mathrm{out}{#2}}^{#1}}\xspace}
\newcommandx{\Mdotout}[2][1=,2=]{\ensuremath{\dot{M}_{\mathrm{out}{#2}}^{#1}}\xspace}
\newcommand\sbullet[1][.5]{\mathbin{\vcenter{\hbox{\scalebox{#1}{$\bullet$}}}}}
\newcommand{\mbh}{\ensuremath{M_{\sbullet[0.85]}}\xspace}

\newcommandx{\fluxdcgs}[1][1=-20]{\ensuremath{\times 10^{#1}~\mathrm{erg\,s^{-1}\,cm^{-2}\,\AA^{-1}}}\xspace}
\newcommandx{\fluxcgs}[2][1=-20,2=\times]{\ensuremath{{#2}10^{#1}~\mathrm{erg\,s^{-1}\,cm^{-2}}}\xspace}
\newcommandx{\powercgs}[2][1=44,2=\times]{\ensuremath{{#2}10^{#1}~\mathrm{erg\,s^{-1}}}\xspace}
\newcommandx{\ergs}{\ensuremath{\mathrm{erg\,s^{-1}}}\xspace}
\newcommand{\AV}{\ensuremath{A_V}\xspace}

\newcommand{\cigale}{{\tt cigale}\xspace}
\newcommand{\jwst}{\textit{JWST}\xspace}
\newcommand{\hst}{\textit{HST}\xspace}
\newcommand{\ppxf}{{\tt ppxf}\xspace}
\newcommand{\eazy}{{\tt eazy}\xspace}
\newcommand{\prospector}{{\tt prospector}\xspace}
\newcommand{\emcee}{{\tt emcee}\xspace}
\newcommand{\cloudy}{{\tt cloudy}\xspace}
\newcommand{\pyneb}{{\tt pyneb}\xspace}
\newcommandx{\mappings}[1][1=]{{\tt mappings{#1}}\xspace}
\newcommand{\galfit}{{\tt galfit}\xspace}
\newcommand{\qubespec}{{\tt qubespec}\xspace}
\newcommand{\pysersic}{{\tt pysersic}\xspace}
\newcommand{\msaexp}{{\tt msaexp}\xspace}
\newcommand{\astropy}{{\tt astropy}\xspace}

\newcommand{\blackthunder}{BlackTHUNDER\xspace}
\newcommand{\Mdynvalue}{$\Mdyn = 2.0\pm0.5 \times 10^{11}$~\MSun}

\defcitealias{gordon+2003}{G03}
\newcommand{\target}{Irony\xspace}
\newcommand{\BIC}{\text{BIC}\xspace}
\newcommand{\sign}{\ensuremath{\sigma_\mathrm{n}}\xspace}
\newcommand{\erica}[1]{\textcolor{green}{#1}}

\begin{document}

\title{Irony at z=6.68: a bright AGN with forbidden Fe emission and multi-component Balmer absorption}

\author[orcid=0000-0003-2388-8172,sname='Francesco D'Eugenio']{Francesco D'Eugenio}
\altaffiliation{Co-first authors}
\affiliation{Kavli Institute for Cosmology, University of Cambridge, Madingley Road, Cambridge, CB3 0HA, UK}
\affiliation{Cavendish Laboratory, University of Cambridge, 19 JJ Thomson Avenue, Cambridge, CB3 0HE, UK}
\email{francesco.deugenio@gmail.com}
\author[orcid=0000-0002-7524-374X,sname='Erica Nelson']{Erica Nelson}
\altaffiliation{Co-first authors}
\affiliation{Department for Astrophysical and Planetary Science, University of Colorado, Boulder, CO 80309, USA}
\email[show]{erica.june.nelson@colorado.edu}
\author[orcid=0000-0002-1660-9502,sname='Xihan Ji']{Xihan Ji}
\affiliation{Kavli Institute for Cosmology, University of Cambridge, Madingley Road, Cambridge, CB3 0HA, UK}
\affiliation{Cavendish Laboratory, University of Cambridge, 19 JJ Thomson Avenue, Cambridge, CB3 0HE, UK}
\email{xj254@cam.ac.uk}
\author[orcid=0009-0005-2295-7246,sname='Josephine Baggen']{Josephine Baggen}
\affiliation{Astronomy Department, Yale University, 219 Prospect St, New Haven, CT 06511, USA}
\email{josephine.baggen@yale.edu}
\author[orcid=0000-0002-5612-3427,sname='Jenny Greene']{Jenny Greene}
\affiliation{Department of Astrophysical Sciences, Princeton University, Princeton, NJ 08544, USA}
\email{jennyg@princeton.edu}
\author[orcid=0000-0002-2057-5376,sname='Ivo Labbe']{Ivo Labb\'e}
\affiliation{Centre for Astrophysics and Supercomputing, Swinburne University of Technology, Melbourne, VIC 3122, Australia}
\email{ilabbe@swin.edu.au}
\author[orcid=0000-0003-0736-7879,sname='Gabriele Pezzulli']{Gabriele Pezzulli}
\affiliation{Kapteyn Astronomical Institute, University of Groningen, Landleven 12, NL-9747 AD Groningen, the Netherlands}
\email{pezzulli@astro.rug.nl}
\author[orcid=0009-0006-5224-3778,sname='Vanessa Brown']{Vanessa Brown}
\affiliation{Department for Astrophysical and Planetary Science, University of Colorado, Boulder, CO 80309, USA}
\email{vanessa.brown@colorado.edu}
\author[orcid=0000-0002-4985-3819,sname='Roberto Maiolino']{Roberto Maiolino}
\affiliation{Kavli Institute for Cosmology, University of Cambridge, Madingley Road, Cambridge, CB3 0HA, UK}
\affiliation{Cavendish Laboratory, University of Cambridge, 19 JJ Thomson Avenue, Cambridge, CB3 0HE, UK}
\affiliation{Department of Physics and Astronomy, University College London, Gower Street, London WC1E 6BT, UK}
\email{rm665@cam.ac.uk}
\author[orcid=0000-0003-2871-127X,sname='Jorryt Matthee']{Jorryt Matthee}
\affiliation{Institute of Science and Technology Austria (ISTA), Am Campus 1, 3400 Klosterneuburg, Austria}
\email{jorryt.matthee@ist.ac.at}
\author[orcid=0000-0000-0000-0000,sname='Elena Terlevich']{Elena Terlevich}
\affiliation{Instituto Nacional de Astrof\'isica, \'Optica y Electr\'onica, 72840 Tonantzintla, Puebla, M\'exico}
\affiliation{Facultad de Astronom\'ia y Geof\'isica, Universidad de La Plata, La Plata, B1900FWA, Argentina}
\email{et@ast.cam.ac.uk}
\author[orcid=0000-0001-6774-3499,sname='Roberto Terlevich']{Roberto Terlevich}
\affiliation{Instituto Nacional de Astrof\'isica, \'Optica y Electr\'onica, 72840 Tonantzintla, Puebla, M\'exico}
\affiliation{Institute of Astronomy, University of Cambridge, Cambridge, CB3 0HA, UK}
\affiliation{Facultad de Astronom\'ia y Geof\'isica, Universidad de La Plata, La Plata, B1900FWA, Argentina}
\email{rjt37@cam.ac.uk}
\author[orcid=0000-0001-5586-6950,sname='Alberto Torralba']{Alberto Torralba}
\affiliation{Institute of Science and Technology Austria (ISTA), Am Campus 1, 3400 Klosterneuburg, Austria}
\email{alberto.torralba@ista.ac.at}
\author[orcid=0000-0002-6719-380X,sname='Stefano Carniani']{Stefano Carniani}
\affiliation{Scuola Normale Superiore, Piazza dei Cavalieri 7, I-56126 Pisa, Italy}
\email{stefano.carniani@sns.it}

\begin{abstract}
We present the deepest medium-resolution JWST/NIRSpec spectroscopy to date of a bright Little Red Dot (LRD) AGN, \textit{Irony} at $z=6.68$. The data reveal broad Balmer emission from \Halpha--\Hdelta and Balmer absorption in \Halpha--\Hepsilon. The absorption lines are kinematically split: H$\alpha$ is blueshifted while higher-order lines are redshifted suggesting complex gas kinematics; their relative ratios are inconsistent with a single, passive absorbing screen. The line depths require absorption of both the BLR and the continuum, ruling out a stellar origin, consistent with the smooth Balmer break. We fit the broad \Hgamma--\Halpha lines and find the data favor a double-Gaussian effective profile, although exponential wings are evident. Depending on the adopted profile, single-epoch virial estimates give $\log(\mbh/\Msun)=7.86\text{--}8.39$ and $\lambda_{\rm Edd}=1.7\text{--}0.4$. The dynamical mass implied by the narrow lines is low $\log(\mdyn/\Msun)=9.1$, suggesting an overmassive black hole. The narrow lines display little attenuation, $\AV<0.5$ mag; while broad $\Halpha/\Hbeta\sim9$ and the broad Balmer decrements are inconsistent with standard dust attenuation curves, suggesting collisional processes. The forbidden-line spectrum includes auroral \SII and \NII, and a forest of \FeII lines. Line ratios and kinematics indicate a stratified narrow-line region with both low ($\nelec=420~\percm$) and high densities ($\nelec\gtrsim6.3\times10^5~\percm$). We detect metal absorption lines in both the optical (\CaII and \NaI) and UV range (\FeIIperm UV1--UV3). Our results support a picture of a compact AGN embedded in a dense, high covering-factor and stratified cocoon, with complex neutral-gas kinematics. While the choice of broad-line profile affects the virial estimates of \mbh, we find the effect to be of order 0.6 dex between the different approaches.
\end{abstract}



\section{Introduction}
JWST has revealed a remarkable population of compact, red objects at $z>4$ that are challenging our understanding of the early coevolution of galaxies and supermassive black holes. These objects, dubbed Little Red Dots (LRDs) by \cite{matthee+2024} are characterized observationally by broad Balmer lines, compact morphology, and V-shaped spectral energy distribution \citep[SED;][]{furtak+2024,killi+2024,greene+2024}.  
Interestingly, the number densities of LRDs seem to drop significantly toward lower redshifts \citep[e.g.][]{kocevski+2024,ma+2025}, suggesting that the physical conditions required for their formation may be much more common at early cosmic times. 
Despite vastly outnumbering UV-selected quasars of similar luminosity at similar redshifts \citep[e.g.][]{harikane+2023,matthee+2024}, the nature of LRDs remains highly enigmatic.

Our physical understanding of LRDs remains limited because the mechanisms responsible for their different emission components are unclear. While their 
broad Balmer lines (FWHM $\gtrsim 1000$ km s$^{-1}$) are reminiscent of 
Type I AGN \citep[e.g.][]{greene+2024}, LRDs display a number of important differences with respect to both local and more luminous AGN. 
These differences include X-ray weakness
\citep{yue+2024,maiolino+2025x}, radio weakness
\citep[][]{mazzolari+2024r,gloudemans+2025}, lack or weakness
of high-ionization lines \citetext{particularly \HeII, \citealp{juodzbalis+2024b,
wang+2025b}; although some high-ionization lines have been observed,
\citealp{killi+2024,tang+2025a,tripodi+2024,labbe+2024,lambrides+2025}}, and emission-line ratios more consistent with photoionization by star formation than an AGN \citep[e.g.,][]{kocevski+2023,ubler+2023,
harikane+2023,maiolino+2024a,juodzbalis+2025,wang+2025b}.
Other anomalies include weakness or complete lack of the prominent, permitted \FeIIperm
humps typical of AGN \citep{trefoloni+2025}, and the low/rare variability, although
detected in some cases \citep{kokubo+2024,zhang+2025,ji+2025a,furtak+2025,naidu+2025}.
Further complicating their 
interpretation, with non-detections in far-infrared observations, their red colors are unlikely to result from dust attenuation 
\citep[e.g.][]{setton+2025}.

As a result of these seemingly contradictory observations, LRDs have found themselves at the center of two opposite controversies. 
If LRDs are interpreted as broad-line AGN, using single-epoch virial calibrations \citep[e.g.][]{greene+ho2005,reines+volonteri2015}, their line widths imply large supermassive black hole masses \citep[\mbh;][]{harikane+2023,maiolino+2024a,furtak+2024, matthee+2024}. Both how to grow supermassive black holes so early and how to explain the high implied \mbh-to-\mstar ratios \citep{harikane+2023,maiolino+2024a}, pose challenges to our theoretical understanding of SMBH growth \citep[e.g.][]{furtak+2024,kokorev+2023,pacucci+2024,dayal+24,schneider+2023,trinca+2024,kritos+2025,rantala+2025}.
Alternatively, if their Balmer breaks and red SEDs reflect 
evolved stellar populations \citep{labbe+2023,Baggen2024,labbe+2024,wang+2024b}, some 
LRDs reach stellar masses that approach or exceed the baryon budget 
available in a $\Lambda$CDM cosmology at such early times 
\citep{boylan-kolchin+2023,inayoshi+ichikawa2024}. Both interpretations seemingly require either exotic 
formation pathways or revisions to our understanding of early cosmic 
evolution.

Even more recent observations have uncovered a set of signatures that rule 
in favor of a non-stellar origin for the observed emission in LRDs. Time 
variability in the equivalent width of the broad lines implies an AGN origin
\citep{ji+2025a,furtak+2025}, while variability in the continuum, 
reduces the stellar mass estimates \citep{ji+2025a,naidu+2025}. 
Clustering analysis finds no evidence for large dark-matter halos, as would 
be expected from objects with large \mstar \citep{pizzati+2024,matthee+2024b,
lin+2025c,arita+2025,zhuang+2025}. Furthermore, Balmer breaks in some 
objects have been found to be too large and too smooth to be explained by 
a stellar population \citep{ji+2025a,naidu+2025,degraaff+2025}. Recent 
observations have also detected line absorption in \HeIL[1.08\mum] 
\citep{wang+2025,juodzbalis+2024b,loiacono+2025} and the Balmer series 
\citep{matthee+2024,juodzbalis+2024b,deugenio+2025d,deugenio+2025e,ma+2025,
naidu+2025,degraaff+2025,lin+2025a}, with depths and widths in the latter
inconsistent with a stellar origin \citep{matthee+2024,juodzbalis+2024b,
deugenio+2025d}. Another key observational evidence disfavoring large
stellar masses are the low dynamical masses inferred from high resolution
spectroscopy \citep[e.g.,][]{ji+2025a,deugenio+2025d,juodzbalis+2025b}.

The Balmer breaks and absorption lines also have another implication. 
Absorption from the unstable $n=2$ level of hydrogen requires very high 
densities and/or very high column densities
of neutral hydrogen \citep[e.g.,][]{inayoshi+maiolino2025,
juodzbalis+2024b,ji+2025a}. This has led to the interpretation of these 
sources as `quasi-stars' \citep{begelman+2006,begelman+dexter2025} or 
`black-hole stars' \citep{naidu+2025}, in which we are observing an early 
super-Eddington growth stage of SMBHs embedded in dense gaseous cocoons. 
This dense gas hypothesis has significant explanatory power for the observed 
properties of LRDs, making it an attractive if surprising explanation. In 
addition to naturally explaining Balmer breaks too strong to be produced by 
stellar populations, dense gas models can account for LRDs' characteristic 
X-ray weakness and red continuum shapes.

The recent discovery of LRDs at low redshifts $z=0.1\text{--}0.3$ \citep{lin+2025b,ji+2025b}
enabled the first detailed look into their absorption features and weak emission lines,
revealing widespread metal absorption, high gas density, and forbidden \FeII emission.
Metal and hydrogen absorption display different kinematics, implying a different origin
\citep[e.g.,][]{ji+2025b}. 
Notably, no known low-redshift LRD displays a clear Balmer break \citetext{cf. \citealp{lin+2025b,ji+2025b} and \citealp{furtak+2024,greene+2024}}, meaning that high-redshift LRDs may still be qualitatively different from their low-redshift counterparts.

Despite these advances, many critical questions remain about the physical properties of LRDs. The properties of Balmer absorption show puzzling inconsistencies that suggest gas kinematics more complex than typically used in models 
\citep{ji+2025a,deugenio+2025d,deugenio+2025e}. A fraction of LRDs also 
exhibit exponential wings in their broad Balmer lines \citep{rusakov+2025}, potentially 
indicating electron scattering in dense environments, with implications for SMBH mass estimates based on broad line widths. Recently, forbidden \FeII emission has been
reported in a LRD at $z=6.68$, together with a higher-ionization \FeVIIL emission
\citep{lambrides+2025}. These detections resonate with local LRD analogues, where
\FeII is clearly detected \citep{lin+2025b,ji+2025b}, and where high-ionization \FeVL
has also been reported \citep{lin+2025b}. The \FeII emission is further supported by
a number of low-resolution prism observations \citetext{\citealp{tripodi+2025};
see also \citealp{taylor+2025b}, their Figs. 1 and~2}.
If confirmed, \FeII and \FeVIIL detection in high-redshift LRDs would imply photoionization and/or density conditions very different
from quasars, where instead we routinely observe permitted \FeIIperm \citep{baldwin+2004,gaskell+2021}.
Moreover, this would cement the local discoveries as fully fledged LRDs, not local `analogs'.
Resolving the questions of the profile shape and iron lines requires deep spectroscopy
capable of detecting weak features, with higher resolution than the prism, required
for deblending ambiguous emission lines and for constraining line 
kinematics.

To address these questions, we present the deepest medium-resolution spectroscopy
of a bright LRD to date, capable of detecting weak emission lines and absorption
features in the continuum. Our target, here called `\target', is a bright LRD at $z=6.68$
\citetext{RUBIES-49140, \citealp{wang+2024b,wang+2025b}; CEERS-10444,
\citealp{kocevksi+2025,tang+2025a,ronayne+2025}; THRILS-46403, \citealp{lambrides+2025}}.
This source was initially selected as a massive-galaxy candidate, but ironically
we find an overmassive black hole; a forest of clearly detected \FeII lines
adds to the iron(y) too.

This work is organized as follows. After describing the observations and data
reduction (Section~\ref{s.data}), we present an updated morphological analysis
(Section~\ref{s.morph}). We then analyze the grating spectrum, focusing on the
broad lines and Balmer absorption (Section~\ref{s.elmodel}) and on the
narrow lines (Section~\ref{s.narrow}). In Section~\ref{sec:cloudy_break} we
use the prism spectrum to inform AGN photo-ionization models.
We discuss the nature of \target in Section~\ref{s.disc} and we conclude with
a summary of our findings (Section~\ref{s.conc}).

Throughout this work, we use a flat \textLambda CDM cosmology with $H_0 = 67.4$~\kms~Mpc$^{-1}$ and $\Omega_\mathrm{m}=0.315$ \citep{planck+2020}, giving a physical scale of 5.49~kpc~arcsec$^{-1}$ at redshift $z=6.68$ (all physical scales are given as proper quantities).
Stellar masses are total stellar mass formed, assuming a \citet{chabrier2003} initial mass function, integrated between 0.1 and 120~\Msun.
All magnitudes are in the AB system \citep{oke+gunn1983} and all equivalent widths (EWs) are
in the rest frame, with positive EW corresponding to line absorption.

\section{Observations and data reduction}\label{s.data}

The spectroscopic data comes from the Cycle-2 \jwst program ID 4106 (PIs E.~J.~Nelson and I.~Labb\'e).
The original observations incurred a `mirror-tilt' event and resulted in unusable grating
exposures; compensation time was awarded in Cycle 3. 
The final observations, which are the basis of this work, consist of a single pointing with
the NIRSpec spectrograph \citep{jakobsen+2022} and the Micro Shutter Assembly \citep[MSA;][]{ferruit+2022}.
We use a single configuration designed for the prism disperser; the same configuration was
also observed with the G395M grating, which incurs spectral overlaps, due to the longer
detector footprint of the gratings compared to the prism. Primary targets incur no spectral
overlap; to achieve this, a few non-primary targets were removed from the MSA configuration in
G395M.

\begin{figure*}
\centering
{\phantomsubcaption\label{f.obs.a}
 \phantomsubcaption\label{f.obs.b}
 \phantomsubcaption\label{f.obs.c}}
\includegraphics[width=\textwidth]{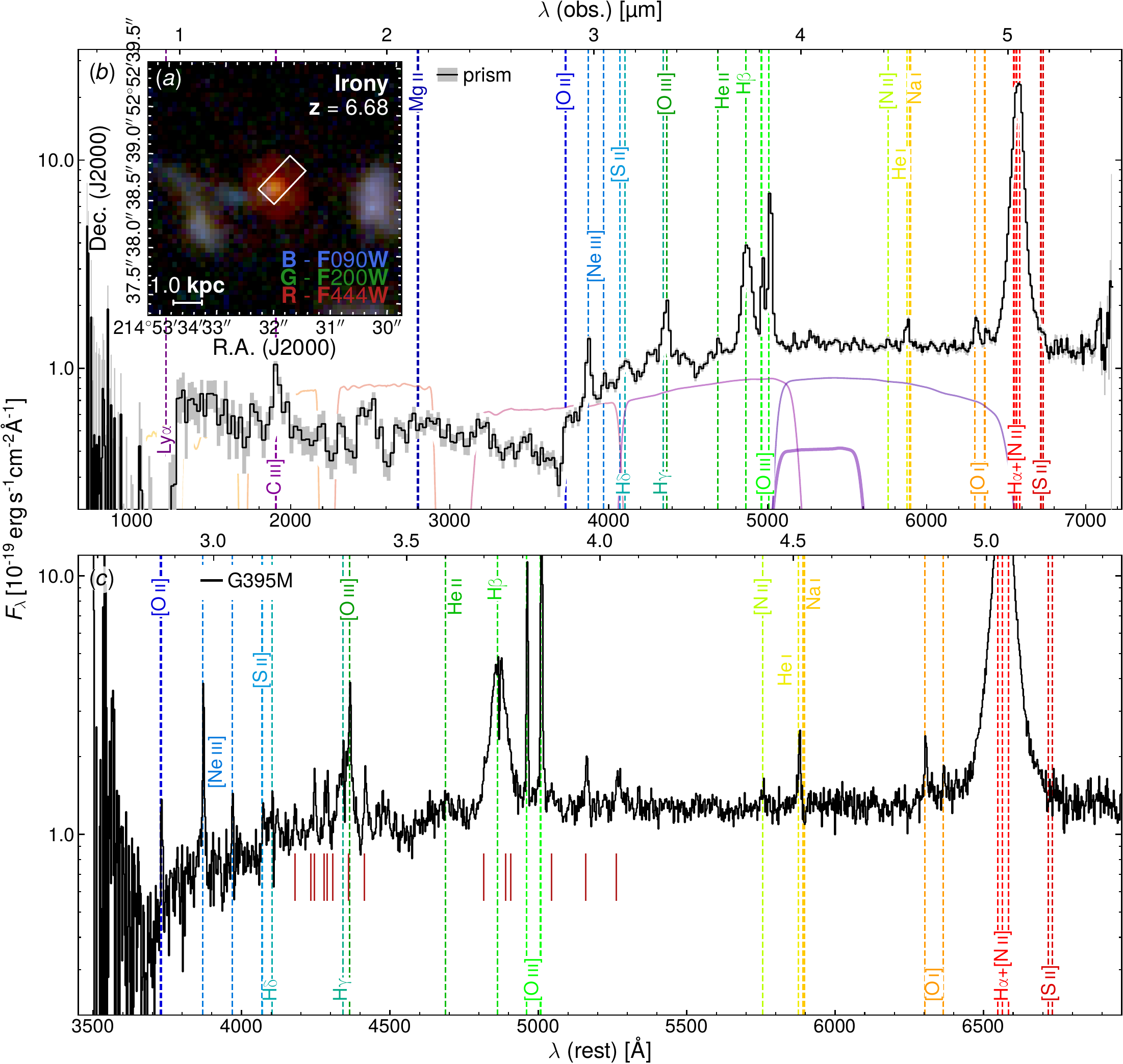}
\caption{Panel~\subref{f.obs.a}. False-color NIRCam image of \target,
with the central MSA shutter overlaid. Panel~\subref{f.obs.b}. Prism spectrum, highlighting the
sharp \Lyalpha break, UV absorption lines, and strong Balmer break; we also show the
transmission of the NIRCam filters available for this object.
Panel~\subref{f.obs.c}. G395M spectrum; the smooth Balmer break rules out an origin in
standard stellar populations. The spectrum displays a rich set of auroral lines, as well as
\FeII lines (solid red markers).
}\label{f.obs}
\end{figure*}

The prism integrations consist of three nodded exposures, each with three 16-groups
integrations, using the \texttt{NRSIRS2} readout \citep{rauscher+2012}. For G395M, we use
two times three nodded exposures, each with three 21-groups integrations. The on-source time
is 10.6 and 26.8~ks for the prism and G395M, respectively.
The data reduction used \msaexp \citep{brammer+2023}, corresponding to version 4
on the \href{https://dawn-cph.github.io/dja}{DAWN JWST Archive}. The reduced and
extracted prism and G395M spectra are shown in Fig.~\ref{f.obs}.
We also use \jwst/NIRCam images from the publicly available survey CEERS
\citep[the Cosmic Evolution Early Release Science Survey;][]{bagley+2023,
finkelstein+2023}.

\section{Morphology}\label{s.morph}

\begin{figure}
    \centering
    \includegraphics[width=\linewidth]{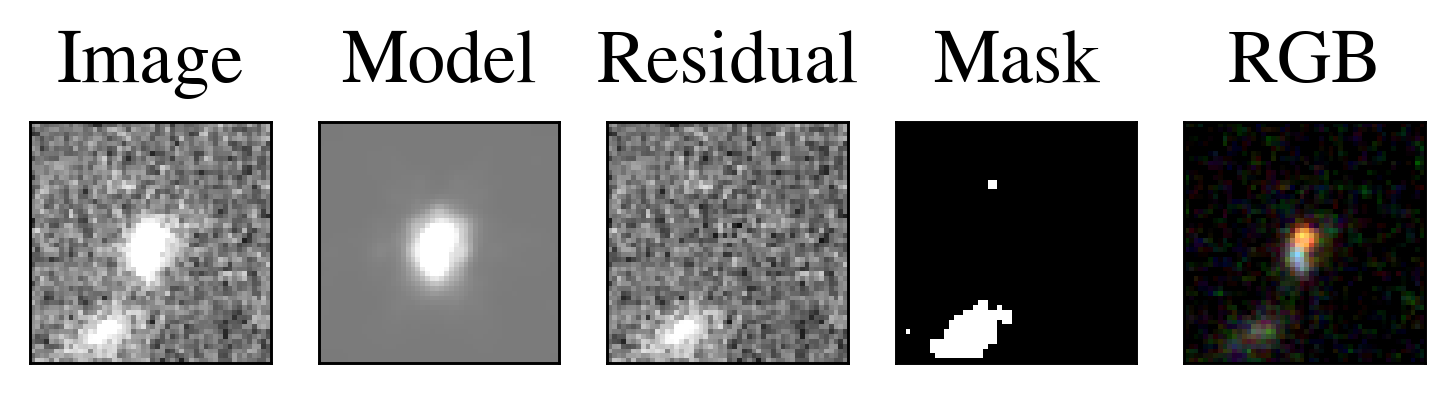}
    \caption{Morphological analysis of \target. From left to right: (1) NIRCam F200W image; (2) best-fitting PSF-convolved \galfit model; (3) residual obtained by subtracting the model from the image; (4) mask;
    (5) RGB image constructed from F115W (blue), F150W (green), and F200W (red). All images are $1\arcsec \times 1\arcsec$.}
    \label{fig:galfit_f200w}
\end{figure}

The morphology of this system was first analyzed in \citet{Baggen2024}, where it was shown to comprise two distinct components: a compact, red central dot and an offset, bluer component. Accordingly, the source is modeled with two \citet{Sersic1968} components in the F200W band, using \galfit \citep{Peng2002, Peng2010} after masking contaminating sources and bad pixels, following the procedures described in \citet{Baggen2024}. The results are shown in Fig.~\ref{fig:galfit_f200w}. 
The best-fit parameters of the red component measured in F200W are $R_{\rm e}$ = 1.4 pixel, S\'ersic index $n=1.77$, $b/a=0.64$, corresponding to a physical circularized effective radius at $z=6.68$ of $R_{\rm e, circ} = R_{\rm e, maj}\sqrt{b/a} = 126 \rm \, pc$.

\begin{figure}
    \centering
    \includegraphics[width=\linewidth]{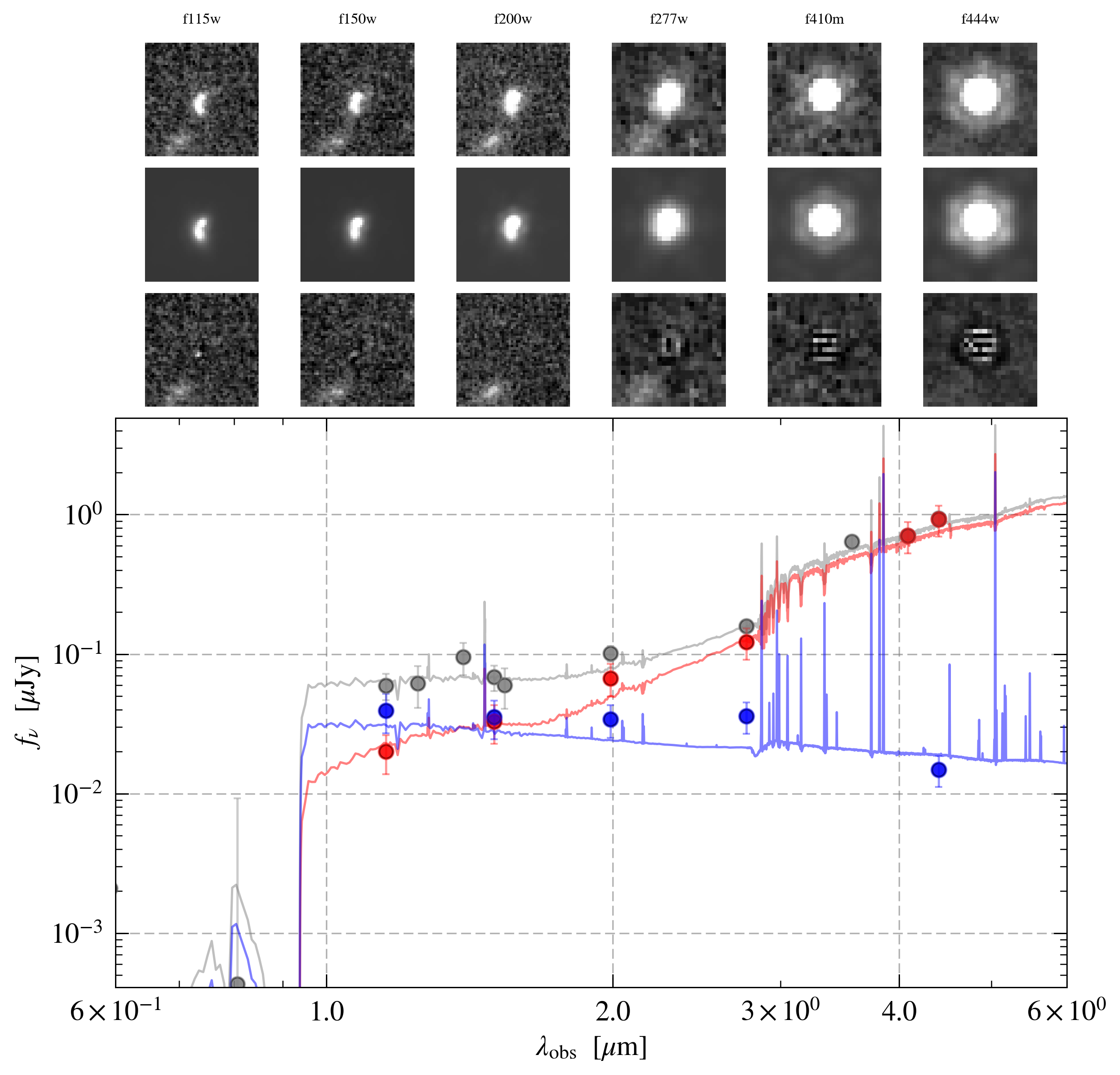}
    \caption{Multi-band \jwst/NIRCam images and SED modeling of \target, where
    the continuum and emission lines are interpreted as due to direct or reprocessed
    stellar emission.
    }\label{fig:morph_sed}
\end{figure}

We perform morphological analysis in all other available bands. These fits start from the F200W results and are propagated to the remaining filters with small allowed variations in the structural parameters, while keeping the magnitudes free (see J.~Baggen (\textit{in prep})). This procedure allows us to measure the flux of both the compact red component and the offset blue component in each filter, producing separate SEDs for the two components.
In F200W, the red and blue components contribute approximately equally to the total flux. As may be expected, the blue component dominates the flux at wavelengths $<2\mu$m, while at wavelengths $>2\mu$m, the galaxy is entirely dominated by the central red component. The total photometry as well as the photometry decomposed into the red and blue components are shown in Fig.~\ref{fig:morph_sed}. 

We also report the results of fitting the red and blue components with stellar population models using \eazy \citep{brammer+2008}. 
The resulting stellar masses are $\log(\mstar/\Msun) = 11.26$ using the total photometry, $\log(\mstar/\Msun) = 11.22$ using the red photometry, and $\log(\mstar/\Msun) = 8.8$ for the blue component. These measurements are consistent with those obtained using \prospector \citep{johnson+2021} by \citet{wang+2024b}. Their \mstar were derived from the spectrum of the entire galaxy combined with total photometry. For further information on separating the blue and red photometry, we refer to J.~Baggen (\textit{in prep.}). Here we present these numbers only for completeness, since they rely on the assumption that both components can be described by stellar populations, but we discuss the nature of the Irony in more detail in Section~\ref{s.disc.ss.mass}. 

LRDs have been selected by their V-shaped SED with a blue color in the rest-UV and red color in the rest-optical. While the integrated photometry of \target abundantly meets the standard selection criteria, this is not the case when the red and blue components are decomposed: the blue color in the rest-UV is driven significantly by the offset blue component. Considering the red component alone yields F150W-F200W=0.77~mag. While this color qualifies this object as an LRD by the criteria of \citet{greene+2024} (F150W-F200W$<0.8$), it no longer qualifies according to the criteria of \citet{kocevski+2024} (F150W-F200W$<0.5$). As will be discussed further in J.~Baggen (\textit{in prep}), spatial decomposition of LRD photometry may have significant implications for our physical interpretation of their SEDs. 

\section{Broad Balmer Lines and Absorption}\label{s.elmodel}
The broad Balmer lines in \target exhibit complex profiles,
with implications for the physical conditions, emission mechanisms, and black hole masses in LRDs. The lines display non-Gaussian shapes 
with exponential-like wings; the choice of broad-line profile affects black hole mass estimates by $\sim$0.6~dex.
Absorption across the Balmer series displays kinematic complexity that is inconsistent with a simple absorbing screen. 
Additionally, the broad Balmer decrements are inconsistent with 
standard dust attenuation curves, ruling out recombination plus dust as an 
explanation and supporting collisional excitation in high-density 
environments. 
We organize our analysis as follows. We describe our modeling approach for 
capturing complex line profiles (\S\ref{s.elmodel.ss.modeling}), describe our 
three alternative broad-line models (\S\ref{s.elmodel.ss.broad}), present 
the model comparison results including their effect on black hole mass 
estimates (\S\ref{s.elmodel.ss.results}), examine the multi-component Balmer 
absorption (\S\ref{s.elmodel.ss.absorption}), and analyze Balmer line ratios 
(\S\ref{s.elmodel.ss.dust}). The reader desiring to skip the modeling details  should proceed to \S\ref{s.elmodel.ss.results}.

\subsection{Modeling approach}\label{s.elmodel.ss.modeling}
To estimate the flux, velocity and width of the emission lines, we use a piecewise fitting approach, following
the method outlined in \citet{deugenio+2025d}. We reduce the number of free parameters
by tying together the properties of emission lines that share similar properties \citep[e.g.,][]{greene+2024}. The model is evaluated on the data grid, after convolving with
the instrument line spread function \citep[LSF;][]{jakobsen+2022}. We use an effective
LSF with higher spectral resolution \citep{greene+2024}, motivated by the fact that
the source is extremely compact \citetext{Section~\ref{s.morph}; \citealp{Baggen2024}}
and does not fill the MSA shutter \citep{degraaff+2024}.

Parameter estimation is performed using a Bayesian approach and a Markov Chain Monte Carlo integrator
\citep[MCMC, \texttt{emcee};][]{foreman-mackey+2013}. 
We use 3,200 walkers and 5,000 steps with 50 percent burn-in, starting chains 
near the global maximum identified using differential evolution 
\citep{storn+price1997}.
Our implementation can be found on \href{https://github.com/fdeugenio/specfitt}{GitHub}.
For model selection, we rely primarily on the Bayes information criterion \citep[\BIC;][]{schwarz1978}, with an
arbitrary threshold of $\Delta \BIC = 10$. 

All narrow emission lines are modeled as
Gaussians, while broad lines use a range of line profiles. Doublets arising from the same
upper level are modeled using the flux of the brightest line as free parameter plus the
fixed flux ratio given by atomic physics \citetext{which we retrieve from \pyneb,
\citealp{luridiana+2015}, unless otherwise specified}.
For any doublet arising from the same lower level, the free parameters are the flux of the bluest
line and the doublet flux ratio, constrained to the range allowed by atomic physics.

\subsection{Broad-line models}\label{s.elmodel.ss.broad}

We model simultaneously \Hgamma, \Hbeta and \Halpha, since these lines are likely to share
the same origin. There are many ways to model the BLR of AGNs \citetext{Gaussian, e.g.,
\citealp{peterson+wandel1999}; logarithmic, e.g., \citealp{blumenthal+mathews1975};
Lorentzian, e.g., \citealp{veron-cetty+2001}; exponential, \citealp{laor2006,rusakov+2025};
and power-law, e.g., \citealp{nagao+2006}}.
In addition, combinations of these models are also possible, and further
emission from broad permitted and forbidden lines is also routinely
observed \citep[e.g.,][]{kakkad+2020}.
However, since LRDs generally display little evidence of ionized outflows
\citep[e.g.,][]{juodzbalis+2024b,deugenio+2025e} and no direct evidence of
broad, permitted \FeIIperm emission in the optical \citep{trefoloni+2025}, we can ignore most of
these complications in our model \citetext{Note that permitted \FeIIperm may be present in the UV, \citealp{labbe+2024,tripodi+2025}}. Below we consider three alternative profiles: exponential, double-Gaussian, and Lorentzian.

\textbf{Exponential model (electron scattering):} 
Following \citet{laor2006} 
and \citet{rusakov+2025}, this model assumes the BLR is a single broad Gaussian 
$\mathcal{G}$ convolved with an exponential kernel $\mathcal{E}$ motivated by 
electron scattering in dense environments:
\begin{equation}\label{eq.expkern}
  \begin{split}
  \mathcal{E}(\lambda) \equiv & \dfrac{1}{2 W} \exp \left( -\left| \dfrac{\lambda - \lambda_0}{W} \right| \right) \\
  W \equiv & \left( 428 \, \tau_\mathrm{e} + 370 \right) \, \dfrac{\lambda_0}{\mathrm{c}} \, \sqrt{\dfrac{T_\mathrm{e}}{10,000~\mathrm{K}}}
  \end{split}
\end{equation}
Where $\lambda_0$ is the observed wavelength of the \Hgamma, \Hbeta or \Halpha, $\tau_\mathrm{e}$
is the wavelength-independent optical depth to electron scattering, $T_\mathrm{e}$ is the
electron temperature, and $\mathrm{c}$ is the speed of light in \kms.
For each broad line, the BLR profile is described by the sum of two components, given by
\begin{equation}
  \exp(-\tau_\mathrm{e}) \, \mathcal{G} + (1 - \exp(-\tau_\mathrm{e})) \, (\mathcal{G} * \mathcal{E}),
\end{equation}
where the first addend is the transmitted and attenuated Gaussian component, while the
second addend is the scattered and broadened component.
The kernel convolution is integrated across 25 times the kernel width $W$,
with the latter linked to $\tau_\mathrm{e}$ and $T_\mathrm{e}$ by the
parameterization proposed in \citet{rusakov+2025}, as shown in Eq.~\ref{eq.expkern}.

\textbf{Double-Gaussian model (empirical):} 
We fit the three Balmer lines simultaneously with two Gaussian components
\citep[e.g.][]{zhu+2009,maiolino+2024a,deugenio+2025d}.
The component centroids
(velocities) are tied across all lines (so only two velocities are free) and the component FWHMs
are likewise tied across lines (two more free parameters). The flux ratio of the two components is
also tied to be identical for all lines (one free parameter). With separate total fluxes for
\Hgamma, \Hbeta and \Halpha, the model has eight free parameters overall.
We remark that the double-Gaussian model does not represent two physically
distinct BLRs \citep[as is sometimes the case; e.g.][]{maiolino+2024a,ubler+2025b},
but just an effective way to capture the shape of the broad emission.

\textbf{Lorentzian model:} We also test a Lorentzian line profile, but using a Voigt
function, which results from the convolution of the intrinsic BLR shape
(assumed to be Lorentzian) with the instrument spectral resolution (assumed
to be Gaussian). In this approach, we have only five free parameters, the
velocity and FWHM of the BLR (common to all three lines), and the flux of each line.

Each of the three models share a number of additional free
parameters, which represent the minimal model components necessary
to satisfactorily reproduce the data.
These are emission from adjacent narrow lines, \Hgamma, \Hbeta and \Halpha, the
\FeIIall, \OIIIall and \NIIall fixed-ratio doublets,
a broad \OIIIall component, and two hydrogen absorbers for each
Balmer line, which are necessary to reproduce the absorption profile (Section~\ref{s.elmodel.ss.absorption}).

The narrow lines \Hgamma, \Hbeta, \Halpha, \OIIIall and \NIIall
are parameterized as Gaussians with the same redshift $z_\mathrm{n}$
and intrinsic dispersion \sign; \OIIIall and \NIIall have fixed
doublet ratios of 0.335 \citep{storey+zeippen2000} and 0.327 \citep{dojcinovic+2023}.
The \FeIIall doublet is parameterized by two Gaussians with fixed
doublet ratio \citep[derived from \textsc{pyneb};][]{luridiana+2015}, with redshift
$z_\mathrm{n}$ (shared with the narrow lines), and with dispersion $\sigma_\mathrm{m}$.
To reduce degeneracies between the narrow Balmer lines
and the Balmer absorption, we parameterize the narrow \Hgamma, \Hbeta and \Halpha
with two parameters, the flux of \Halpha and the $V$-band dust attenuation
$A_V$. We force the intrinsic narrow Balmer-line ratios to follow the values
for Case-B recombination and standard temperature and density $T_\mathrm{e} =
10,000$~K, $n_\mathrm{e}=100~\percm[3]$ \citep{osterbrock+ferland2006}. The
observed line fluxes are then calculated by applying the
\citet[][hereafter: \citetalias{gordon+2003}]{gordon+2003} SMC dust-extinction law.

The \OIIIall broad component also has a fixed ratio, but  a separate free velocity $v_\mathrm{out}$ and velocity dispersion $\sigma_\mathrm{out}$,
with $v_\mathrm{out}$ measured relative to the narrow lines, and where we force
$\sigma_\mathrm{out}>\sign$ using an erfc prior.

The two absorbers are modeled with the standard attenuation model \citep[e.g.,][]{juodzbalis+2024b}, namely
\begin{equation}\label{eq.residual}
\begin{split}
    I(\lambda)/I_0(\lambda) &= 1 - C_f + C_f \cdot \exp \left(- \tau(k;\,\lambda) \right)\\
    \tau(k;\,\lambda) &= \tau_0(k) \cdot f[v(\lambda)],
\end{split}
\end{equation}
where $I_0(\lambda)$ is the spectral flux density before absorption, $\tau_0(k)$ is the optical depth at the center of the line (with $k=\Hgamma$, \Hbeta or \Halpha) and $f[v(\lambda)]$ is the velocity distribution of the absorbing atoms, which we take to
be Gaussian.
For each of the two absorbers, there are three optical depth values at line center, i.e. $\tau_1(\Hgamma)$, $\tau_1(\Hbeta)$ and $\tau_1(\Halpha)$ for the first absorber and
similarly $\tau_2(\Hgamma)$, $\tau_2(\Hbeta)$ and $\tau_2(\Halpha)$ for the second absorber. There are two covering factors, $C_\mathrm{f,1}$ and $C_\mathrm{f,2}$, two velocities $v_\mathrm{abs,1}$ and $v_\mathrm{abs,2}$,
and two velocity dispersions $\sigma_\mathrm{abs,1}$ and $\sigma_\mathrm{abs,2}$; these
parameters are shared between all Balmer lines for the same absorber. In total, there
are twelve free parameters for two absorbing clouds across three emission lines.

\subsection{Broad line model comparison}\label{s.elmodel.ss.results}

\begin{table*}
\centering
  \begin{tabular}{l|lr|rrr}
\hline
 & Parameter                                        & Unit                                            &  Exponential              & Double Gaussian           & Lorentzian                \\
\hline                                                                                                                                                                                    
 & BIC                                              & [---]                                           & $-356$                    & $-392$                    & $465$                     \\
\multirow{10}{*}{\rotatebox[origin=c]{90}{Narrow lines}}                                                                                                                                  
 & $z_\mathrm{n}$                                   & [---]                                           &       $6.68481\pm0.00001$ &       $6.68482\pm0.00001$ &       $6.68482\pm0.00002$ \\
 & $\sigma_\mathrm{n}$                              & $[\mathrm{km\,s^{-1}}]$                         &            $55^{+1}_{-1}$ &            $55^{+1}_{-1}$ &            $55^{+1}_{-1}$ \\
 & $\sigma_\mathrm{m}$                              & $[\mathrm{km\,s^{-1}}]$                         &         $320^{+30}_{-20}$ &         $310^{+20}_{-20}$ &         $320^{+30}_{-20}$ \\
 & $A_V$                                            & $[\mathrm{mag}]$                                &    $0.47^{+0.08}_{-0.09}$ &    $0.51^{+0.08}_{-0.09}$ &    $0.07^{+0.06}_{-0.05}$ \\
 & $F_\mathrm{n}(\mathrm{H\gamma})$                 & $[10^{-18} \, \mathrm{erg\,s^{-1}\,cm^{-2}}]$   &    $0.46^{+0.04}_{-0.04}$ &    $0.46^{+0.04}_{-0.04}$ &    $0.61^{+0.03}_{-0.03}$ \\
 & $F(\mathrm{[O\,III]\lambda 4363})$               & $[10^{-18} \, \mathrm{erg\,s^{-1}\,cm^{-2}}]$   &    $0.61^{+0.03}_{-0.03}$ &    $0.62^{+0.03}_{-0.04}$ &    $0.57^{+0.03}_{-0.03}$ \\
 & $F_\mathrm{n}(\mathrm{H\beta})$                  & $[10^{-18} \, \mathrm{erg\,s^{-1}\,cm^{-2}}]$   &    $1.07^{+0.08}_{-0.08}$ &    $1.08^{+0.08}_{-0.09}$ &    $1.33^{+0.06}_{-0.06}$ \\
 & $F(\mathrm{[O\,III]\lambda 5007})$               & $[10^{-18} \, \mathrm{erg\,s^{-1}\,cm^{-2}}]$   &   $10.99^{+0.08}_{-0.09}$ &   $11.02^{+0.08}_{-0.08}$ &   $11.03^{+0.08}_{-0.09}$ \\
 & $F_\mathrm{n}(\mathrm{H\alpha})$                 & $[10^{-18} \, \mathrm{erg\,s^{-1}\,cm^{-2}}]$   &       $3.7^{+0.3}_{-0.3}$ &       $3.7^{+0.3}_{-0.3}$ &       $3.9^{+0.2}_{-0.2}$ \\
 & $F(\mathrm{[N\,II]\lambda 6583})$                & $[10^{-18} \, \mathrm{erg\,s^{-1}\,cm^{-2}}]$   &    $0.16^{+0.08}_{-0.08}$ &    $0.04^{+0.05}_{-0.03}$ &    $0.22^{+0.08}_{-0.08}$ \\
\hline                                                                                                                                                                                    
\multirow{12}{*}{\rotatebox[origin=c]{90}{Broad lines}}
 & $v_\mathrm{BLR}$                                 & $[\mathrm{km\,s^{-1}}]$                         &             $5^{+3}_{-3}$ &            ---            &             $3^{+2}_{-2}$ \\
 & $v_\mathrm{BLR,1}$                               & $[\mathrm{km\,s^{-1}}]$                         &            ---            &            $40^{+4}_{-4}$ &            ---            \\
 & $v_\mathrm{BLR,2}$                               & $[\mathrm{km\,s^{-1}}]$                         &            ---            &           $-91^{+8}_{-7}$ &            ---            \\
 & $FWHM_\mathrm{BLR,1}$                            & $[\mathrm{km\,s^{-1}}]$                         &            ---            &        $1780^{+20}_{-20}$ &            ---            \\
 & $FWHM_\mathrm{BLR,2}$                            & $[\mathrm{km\,s^{-1}}]$                         &            ---            &        $4130^{+30}_{-30}$ &            ---            \\
 & $FWHM_\mathrm{BLR}$                              & $[\mathrm{km\,s^{-1}}]$                         &        $1350^{+50}_{-50}$ &        $2220^{+20}_{-10}$ &        $2580^{+20}_{-20}$ \\
 & $F_\mathrm{BLR,1}/F_\mathrm{BLR}$                & [---]                                           &            ---            & $0.46^{+0.01}_{-0.01}$    &            ---            \\
 & $F_\mathrm{b}(\mathrm{H\gamma})$                 & $[10^{-18} \, \mathrm{erg\,s^{-1}\,cm^{-2}}]$   &       $2.8^{+0.1}_{-0.1}$ &       $2.7^{+0.1}_{-0.1}$ &       $5.3^{+0.3}_{-0.3}$ \\
 & $F_\mathrm{b}(\mathrm{H\beta})$                  & $[10^{-18} \, \mathrm{erg\,s^{-1}\,cm^{-2}}]$   &      $15.5^{+0.2}_{-0.2}$ &      $15.2^{+0.2}_{-0.2}$ &      $24.9^{+0.3}_{-0.3}$ \\
 & $F_\mathrm{b}(\mathrm{H\alpha})$                 & $[10^{-18} \, \mathrm{erg\,s^{-1}\,cm^{-2}}]$   &     $141.2^{+0.6}_{-0.6}$ &     $138.4^{+0.5}_{-0.5}$ &     $178.3^{+0.7}_{-0.7}$ \\
 & $\tau_\mathrm{e}$                                & [---]                                           &       $2.1^{+0.1}_{-0.1}$ &            ---            &            ---            \\
 & $\Telec$                                         & $[10^4\,\mathrm{K}]$                            &    $0.66^{+0.07}_{-0.06}$ &            ---            &            ---            \\
\hline                                                                                                                                                                                    
\multirow{12}{*}{\rotatebox[origin=c]{90}{Hydrogen absorbers}}
 & $v_\mathrm{abs,1}$                               & $[\mathrm{km\,s^{-1}}]$                         &           $-49^{+4}_{-6}$ &           $-46^{+4}_{-5}$ &           $-37^{+2}_{-2}$ \\
 & $\sigma_\mathrm{abs,1}$                          & $[\mathrm{km\,s^{-1}}]$                         &           $116^{+5}_{-5}$ &           $106^{+5}_{-5}$ &           $190^{+5}_{-5}$ \\
 & $C_{f,1}$                                        & [---]                                           &    $0.63^{+0.03}_{-0.03}$ &    $0.59^{+0.03}_{-0.03}$ &    $0.98^{+0.01}_{-0.03}$ \\
 & $\tau_\mathrm{H\gamma,1}$                        & [---]                                           &       $0.9^{+0.2}_{-0.2}$ &       $1.0^{+0.2}_{-0.2}$ &    $0.16^{+0.07}_{-0.05}$ \\
 & $\tau_\mathrm{H\beta,1}$                         & [---]                                           &       $1.8^{+0.3}_{-0.2}$ &       $2.0^{+0.3}_{-0.3}$ &    $0.16^{+0.04}_{-0.04}$ \\
 & $\tau_\mathrm{H\alpha,1}$                        & [---]                                           &       $2.4^{+0.3}_{-0.3}$ &       $2.8^{+0.4}_{-0.3}$ &    $1.12^{+0.06}_{-0.04}$ \\
 & $v_\mathrm{abs,2}$                               & $[\mathrm{km\,s^{-1}}]$                         &         $160^{+10}_{-10}$ &         $160^{+10}_{-10}$ &            $77^{+4}_{-4}$ \\
 & $\sigma_\mathrm{abs,2}$                          & $[\mathrm{km\,s^{-1}}]$                         &          $80^{+10}_{-11}$ &          $80^{+10}_{-10}$ &          $20^{+20}_{-10}$ \\
 & $C_{f,2}$                                        & [---]                                           &    $0.87^{+0.08}_{-0.08}$ &    $0.86^{+0.08}_{-0.08}$ &    $0.80^{+0.04}_{-0.04}$ \\
 & $\tau_\mathrm{H\gamma,2}$                        & [---]                                           &       $1.4^{+0.3}_{-0.2}$ &       $1.4^{+0.3}_{-0.2}$ &            $10^{+3}_{-3}$ \\
 & $\tau_\mathrm{H\beta,2}$                         & [---]                                           &       $1.4^{+0.2}_{-0.2}$ &       $1.5^{+0.2}_{-0.2}$ &       $4.6^{+0.7}_{-0.8}$ \\
 & $\tau_\mathrm{H\alpha,2}$                        & [---]                                           &    $0.13^{+0.05}_{-0.02}$ &    $0.13^{+0.05}_{-0.03}$ &    $0.14^{+0.03}_{-0.02}$ \\
\hline                                                                                                                                                                                    
\multirow{3}{*}{\rotatebox[origin=c]{90}{Outflow}}
 & $F_\mathrm{out}(\mathrm{[O\,III]\lambda 5007})$  & $[10^{-18} \, \mathrm{erg\,s^{-1}\,cm^{-2}}]$   &    $1.23^{+0.08}_{-0.08}$ &    $1.23^{+0.08}_{-0.07}$ &    $1.40^{+0.08}_{-0.08}$ \\
 & $v_\mathrm{out}$                                 & $[\mathrm{km\,s^{-1}}]$                         &         $-40^{+20}_{-20}$ &         $-50^{+30}_{-30}$ &         $-20^{+30}_{-30}$ \\
 & $\sigma_\mathrm{out}$                            & $[\mathrm{km\,s^{-1}}]$                         &         $460^{+40}_{-40}$ &         $470^{+40}_{-40}$ &         $540^{+60}_{-50}$ \\
\hline                                                                                                                                                                                    
\multirow{10}{*}{\rotatebox[origin=c]{90}{Derived Parameters}}
 & $\mathrm{EW(H\gamma,1)}$                         & [\AA]                                           &       $2.1^{+0.3}_{-0.3}$ &       $3.5^{+0.6}_{-0.6}$ &       $1.1^{+0.5}_{-0.3}$ \\
 & $\mathrm{EW(H\beta,1)}$                          & [\AA]                                           &       $4.0^{+0.2}_{-0.2}$ &       $6.5^{+0.5}_{-0.4}$ &       $1.3^{+0.3}_{-0.3}$ \\
 & $\mathrm{EW(H\alpha,1)}$                         & [\AA]                                           &       $5.8^{+0.2}_{-0.2}$ &       $9.3^{+0.3}_{-0.4}$ &       $8.6^{+0.1}_{-0.1}$ \\
 & $\mathrm{EW(H\gamma,2)}$                         & [\AA]                                           &       $3.2^{+0.4}_{-0.4}$ &       $3.7^{+0.4}_{-0.4}$ &       $4.7^{+0.3}_{-0.3}$ \\
 & $\mathrm{EW(H\beta,2)}$                          & [\AA]                                           &       $4.1^{+0.3}_{-0.3}$ &       $4.7^{+0.4}_{-0.4}$ &       $5.1^{+0.2}_{-0.2}$ \\
 & $\mathrm{EW(H\alpha,2)}$                         & [\AA]                                           &       $0.6^{+0.2}_{-0.1}$ &       $0.7^{+0.2}_{-0.1}$ &    $0.42^{+0.09}_{-0.07}$ \\
 & $\log\,SFR(\mathrm{H\alpha})$                    & $[\mathrm{M_\odot\,yr^{-1}}]$                   &    $0.77^{+0.04}_{-0.05}$ &    $0.79^{+0.04}_{-0.04}$ &    $0.67^{+0.03}_{-0.03}$ \\
 & $\log\,L_\mathrm{b}(\mathrm{H\alpha})$           & $[\mathrm{10^{42}\,erg\,s^{-1}}]$               &    $2.03^{+0.02}_{-0.03}$ &    $2.19^{+0.05}_{-0.05}$ &    $2.00^{+0.02}_{-0.01}$ \\
 & $\log\,(M_\bullet)$                              & $[\mathrm{M_\odot}]$                            &    $7.82^{+0.03}_{-0.04}$ &    $8.34^{+0.02}_{-0.03}$ & $8.389^{+0.010}_{-0.009}$ \\
 & $\lambda_\mathrm{Edd}$                           & [---]                                           &       $1.7^{+0.1}_{-0.1}$ &    $0.73^{+0.05}_{-0.05}$ & $0.425^{+0.011}_{-0.010}$ \\
 & $W$                                              & $\mathrm{[km\,s^{-1}]}$                         &        $1032^{+10}_{-10}$ &            ---            &            ---            \\
  \end{tabular}
  \caption{Summary of the broad-line model, including ancillary narrow-line parameters, the hydrogen absorbers, and an ionized outflow traced by broad \OIIIall.
  The last group of parameters are derived from the free parameters, based on the MCMC chains. All lines fluxes are observed, without dust-attenuation correction.
  $^\ddag$ The BLR FWHM for the double Gaussian model is also a
  derived parameter, obtained from the FWHM of the two Gaussians $FWHM_\mathrm{BLR,1}$ and $FWHM_\mathrm{BLR,2}$, and from the Gaussian flux ratio
  $F_\mathrm{BLR,1}/F_\mathrm{BLR}$.
    }
  \label{t.broad}
\end{table*}

\begin{figure*}[!t]
    \centering
    {\phantomsubcaption\label{f.fit.a}
     \phantomsubcaption\label{f.fit.b}
     \phantomsubcaption\label{f.fit.c}
     \phantomsubcaption\label{f.fit.d}
     \phantomsubcaption\label{f.fit.e}
     \phantomsubcaption\label{f.fit.f}
     \phantomsubcaption\label{f.fit.g}
     \phantomsubcaption\label{f.fit.h}
     \phantomsubcaption\label{f.fit.i}}
    \includegraphics[width=\textwidth]{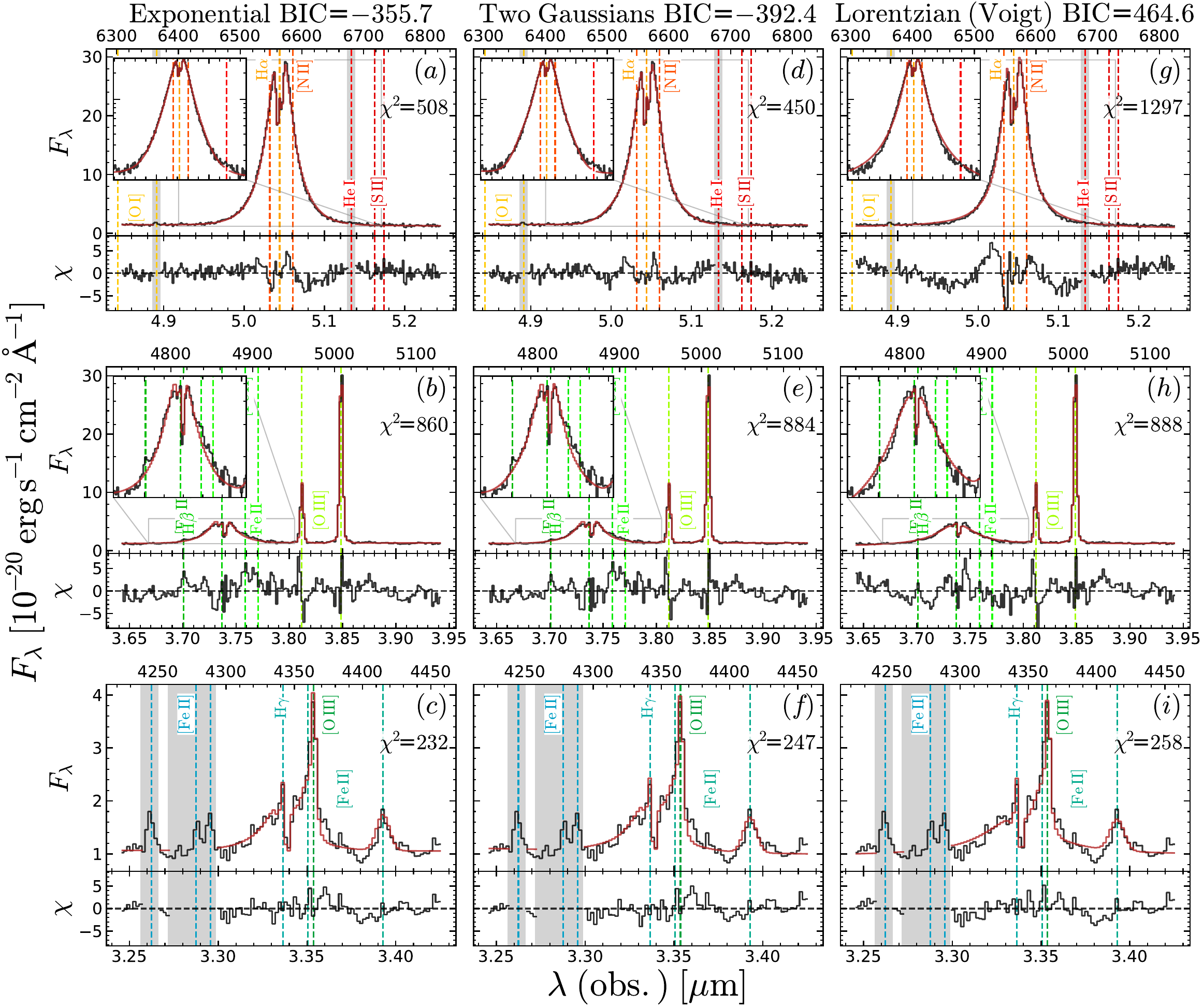}
    \caption{Comparison of the exponential, double-Gaussian and
    Lorentzian models for the BLR of \target, showing the
    \Hgamma--\FeIIall-\OIIIL[4363] complex (top row), the
    \Hbeta--\OIIIall spectral region (middle), and \Halpha
    (bottom); below each spectral region, we also show the
    fit residuals $\chi$. The inset panels show the BLR emission in
    with a logarithmic y scale. The vertical gray regions are
    not fit. In each panel, $\chi^2$ is calculated in the wavelength range fit. Overall, a Lorentzian does not describe
    the data adequately (right column); the exponential and
    double-Gaussian models perform similarly well for \Hgamma
    and \Hbeta, but \Halpha is best described
    by the double-Gaussian model. Notice the different profiles of
    the Balmer absorption, with \Hgamma and \Hbeta appearing
    redshifted, while \Halpha appears blueshifted (Fig.~\ref{f.absorbers}). The iron
    lines have a complex profile, with a blueshifted core and
    a fainter, redshifted wing (panels~\subref{f.fit.c}, \subref{f.fit.f} and~\subref{f.fit.i}).
    }\label{f.fit}
\end{figure*}

The model comparison results are shown in Figure~\ref{f.fit} and 
Table~\ref{t.broad}.
The fit quality (or lack thereof) is quantified by the BIC score.
The complexity of the line profiles can be
appreciated by inspecting the residuals $\chi$, defined as (data - model) /
noise.
The $\Delta \BIC$ overwhelmingly favors the double-Gaussian 
model (BIC = $-392.4$), with the exponential model in second position 
(BIC = $-355.7$, $\Delta \BIC = 37$) and the Lorentzian model strongly ruled 
out (BIC = $464.6$, $\Delta \BIC \gg 10$).

Since the signal-to-noise ratio (SNR) is dominated by \Halpha, we attempt to
separate the contribution of each Balmer line to the result, by calculating
$\chi^2$ over each spectral interval (top right of each panel). Of course,
these `partial' $\chi^2$ do not take into account the different number of
degrees of freedom of the three models. Still, they suggest that the
Lorentzian model performs poorly for all three lines individually.
The Exponential and double-Gaussian models are comparable for \Hgamma and
\Hbeta, with a slight preference for the Exponential model. The difference is
too small to draw strong conclusions, since there are also clear
systematics, such as a spectral dip near 4,400 rest-frame \AA for \Hgamma and
three--four weak emission lines in the wings of \Hbeta. For \Halpha, however, the
double-Gaussian model appears to be superior.
While even the \Halpha fit is affected by systematics (cf. residual sub-panels of
Fig.~\ref{f.fit.c} and~\subref{f.fit.f}) which could skew the results somewhat,
the double-Gaussian model captures better the line profile, leaving smaller residuals
between 4.95 and 5.1 observed-frame \mum. Thus the overall preference for the
double-Gaussian is driven primarily by its better performance at \Halpha, even though
visual inspection shows remarkably exponential line wings (inset panels in
Fig~\ref{f.fit}). This finding is different from the recent work of
\citet{brazzini+2025b}, who studied the \textit{Rosetta Stone} LRD at $z=2.26$
finding that \Halpha alone strongly favors an Exponential model.\footnote{
\citet{brazzini+2025b} reject the scattering due to a screen of free electron based
on other considerations.} In our case, we trace the preference of the double Gaussian over
the Exponential to weak line asymmetries. In the current double-Gaussian model, the
velocity centroids of the two Gaussians differ by $130\pm9~\kms$ (cf. $v_\mathrm{BLR,1}$
and $v_\mathrm{BLR,2}$ in Table~\ref{t.broad}). When we force the same centroid, the
BIC increases to $-273$, tipping the statistical scale in favor of the Exponential.

\begin{figure}[!t]
    \centering
    {\phantomsubcaption\label{f.dustatt.a}
     \phantomsubcaption\label{f.dustatt.b}}
    \includegraphics[width=\columnwidth]{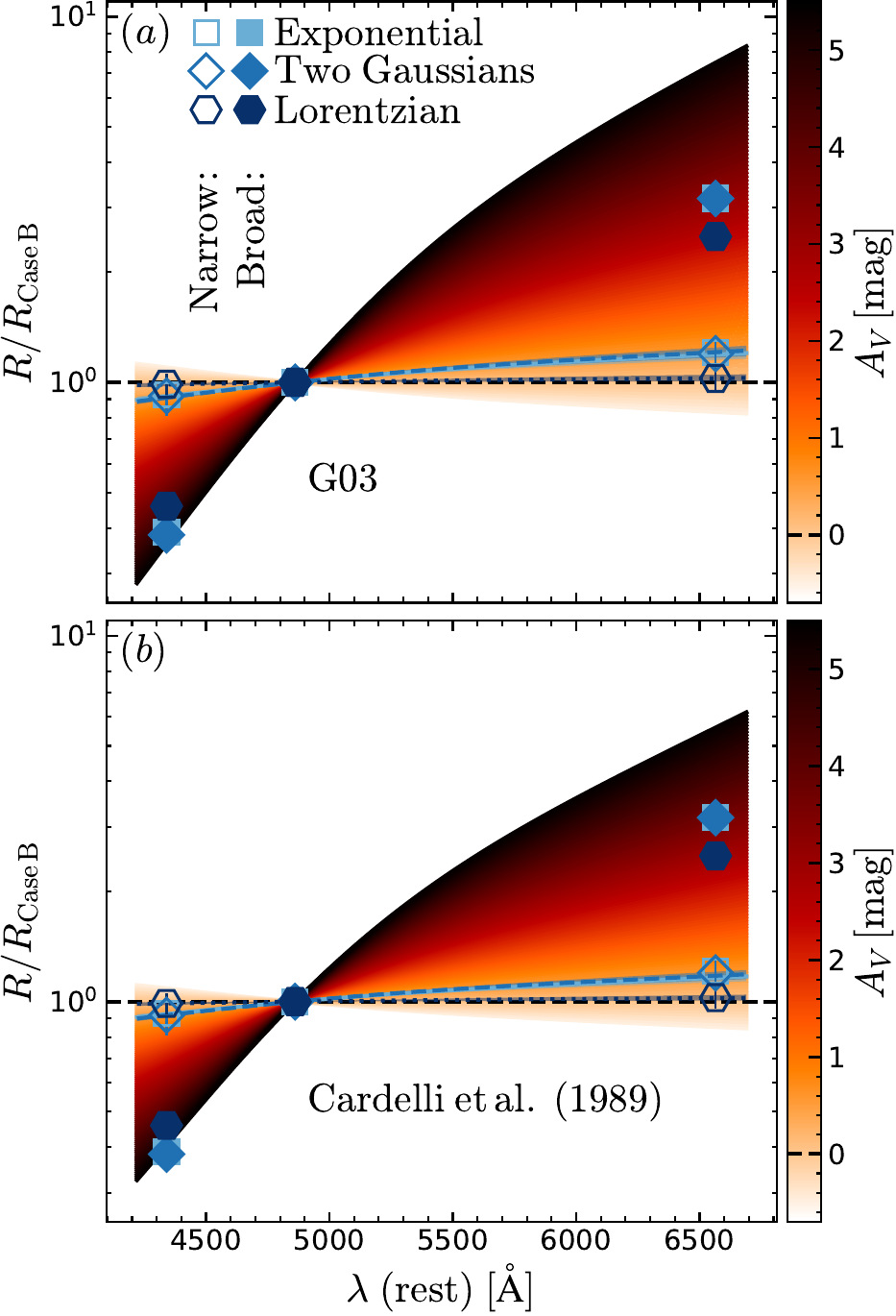}
    \caption{By observing three broad Balmer lines, we can infer that no
    amount of dust (assuming standard attenuation laws) can reconcile the
    line ratios with standard Case-B recombination fluxes. We plot the
    the observed flux ratios \Hgamma/\Hbeta and \Halpha/\Hbeta (shown in the plot at the wavelengths of \Hgamma and \Halpha, respectively), relative to
    the intrinsic ratios from Case-B recombination,
    $R/R_\mathrm{Case\,B}$,. We find that the narrow lines (empty symbols) can
    be explained by a single attenuation curve, while the broad lines yield
    inconsistent attenuation levels from \Hgamma/\Hbeta to \Halpha/\Hbeta
    (solid symbols). Different attenuation laws
    \citep[e.g.][panel~\subref{f.dustatt.b}]{cardelli+1989} can reduce but not
    eliminate the discrepancy.
    }\label{f.dustatt}
\end{figure}

\subsection{Physically interpreting the broad lines}\label{s.elmodel.ss.other}

Crucially, the broad lines have different FWHM, which has
an impact on the black-hole mass measurement. The derived broad line FWHM values Exponential, double-Gaussian
and Lorentzian models infer FWHM values of 1390, 2250, and 2580~\kms,
respectively. These values have random uncertainties of order 15--50~\kms, hence
the differences are highly significant.
Applying the single-epoch virial calibration of \citep{reines+volonteri2015},
we get values of $\log(\mbh/\Msun) = 7.86$, $8.34$, and $8.39$, respectively.
These values decrease to 7.68, 8.21 and 8.25 when using \citet{greene+ho2005}.
The random uncertainties about our \mbh are negligible (0.01~dex) compared to
the scatter about the adopted calibration \citetext{0.3~dex,
\citealp{reines+volonteri2015}; similar values are found about other calibrations,
\citealp{greene+ho2005}; while calibrations based on the observed second moment
of the broad line yield 0.2~dex \citealp{dallabonta+2025}}.

An exponential
model for \target has also been presented in \citet{rusakov+2025}. While we find
$\tau_\mathrm{e} = 2.1$ (Table~\ref{t.broad}), implying that 12\% of the underlying
Gaussian line is not scattered, they do not detect this transmitted Gaussian component,
resulting in an upper limit $\log(\mbh/\Msun)<6.4$. However, their analysis is limited to the \Halpha line
alone, and on much lower-SNR data. This limits their model's ability to
disentangle the absorber, narrow line, and transmitted broad line. They tested
that with the deeper data used here their results become consistent with ours
(V. Rusakov and G. Panagiotis Nikopoulos, priv.~comm.). Further, repeating the exponential
fit while forcing the intrinsic Gaussian to have FWHM$<1,000$~\kms yields a considerably
worse fit relative to the unconstrained exponential model $\Delta\,\BIC>40$.
Having thus ruled out extremely narrow or undetected intrinsic Gaussian even for the
Exponential model, we estimate a systematic discrepancy on \mbh of a factor of four,
depending on the adopted line profile. This propagates to a similar uncertainty
in the Eddington ratios $\lambda_\mathrm{Edd}$, which range from 1.6 to 0.4.

As for the physical interpretation of the models, we treat the double-Gaussian
model as a phenomenological model rather than evidence for two distinct BLRs (or BLR
sub-regions). Our model returns a flux ratio between the narrower broad Gaussian and
the total broad profile of $F_\mathrm{b,1}/F_\mathrm{b} = 0.46\pm0.01$
(Table~\ref{t.broad}; the tied double Gaussian gives $F_\mathrm{b,1}/F_\mathrm{b} =
0.48\pm0.01$). These values match those reported for other LRDs with similar
measurements \citep{deugenio+2025d,deugenio+2025e}. In our view, such uniformity
disfavors a physical interpretation of the two Gaussians: it would imply a nearly
universal flux ratio, contrary to expectations for distinct regions (e.g. two SMBHs).
Indeed, when there is spatial evidence for multiple SMBHs, the flux ratio can differ
markedly from $\approx 0.5$ \citep{ubler+2025b}, and in bright quasars the two
Gaussians often have different velocities \citep{wills+1993,brotherton+1994} with
object-to-object and epoch-to-epoch variations in flux ratio
\citep{marziani+2018,sulentic+2000}. However, we caution that in one of the
objects where the flux ratio is $\approx 0.5$, the narrowest broad component is
found to be spatially extended, unlike the broadest component; this implies a
separate physical nature of the two Gaussians \citep{juodzbalis+2025b}. Clearly,
current samples are still too small, but a universal flux ratio while the two
Gaussians have different origin requires some regulating mechanism.
We do not yet have a physical explanation for this convergence, and flag it as
an open issue.

The Exponential model yields a large optical depth $\tau_\mathrm{e} = 2.0\pm0.1$,
similar to the results of \citet{rusakov+2025}. Together with the line width, the
optical depth constrains the electron temperature of the putative medium embedding
the BLR, which is found to be relatively low: $T_\mathrm{e} = 7,000\pm800$~K. The latter arises due to the combination of
high $\tau_\mathrm{e}$ (which broadens the exponential wings) but relatively
narrow exponential profile, which in the simple model we adopted can only be
reconciled by decreasing the mean energy of the scattering electrons.

We also test a model where there is a single absorber, but there are multiple
narrow-line emitters creating differential line infill. This is discussed in
Section~\ref{s.elmodel.ss.absorption} below.

\subsection{Balmer absorption}\label{s.elmodel.ss.absorption}

\begin{figure}[!t]
    \centering
    {\phantomsubcaption\label{f.absorbers.a}
     \phantomsubcaption\label{f.absorbers.b}
     \phantomsubcaption\label{f.absorbers.c}}
    \includegraphics[width=\columnwidth]{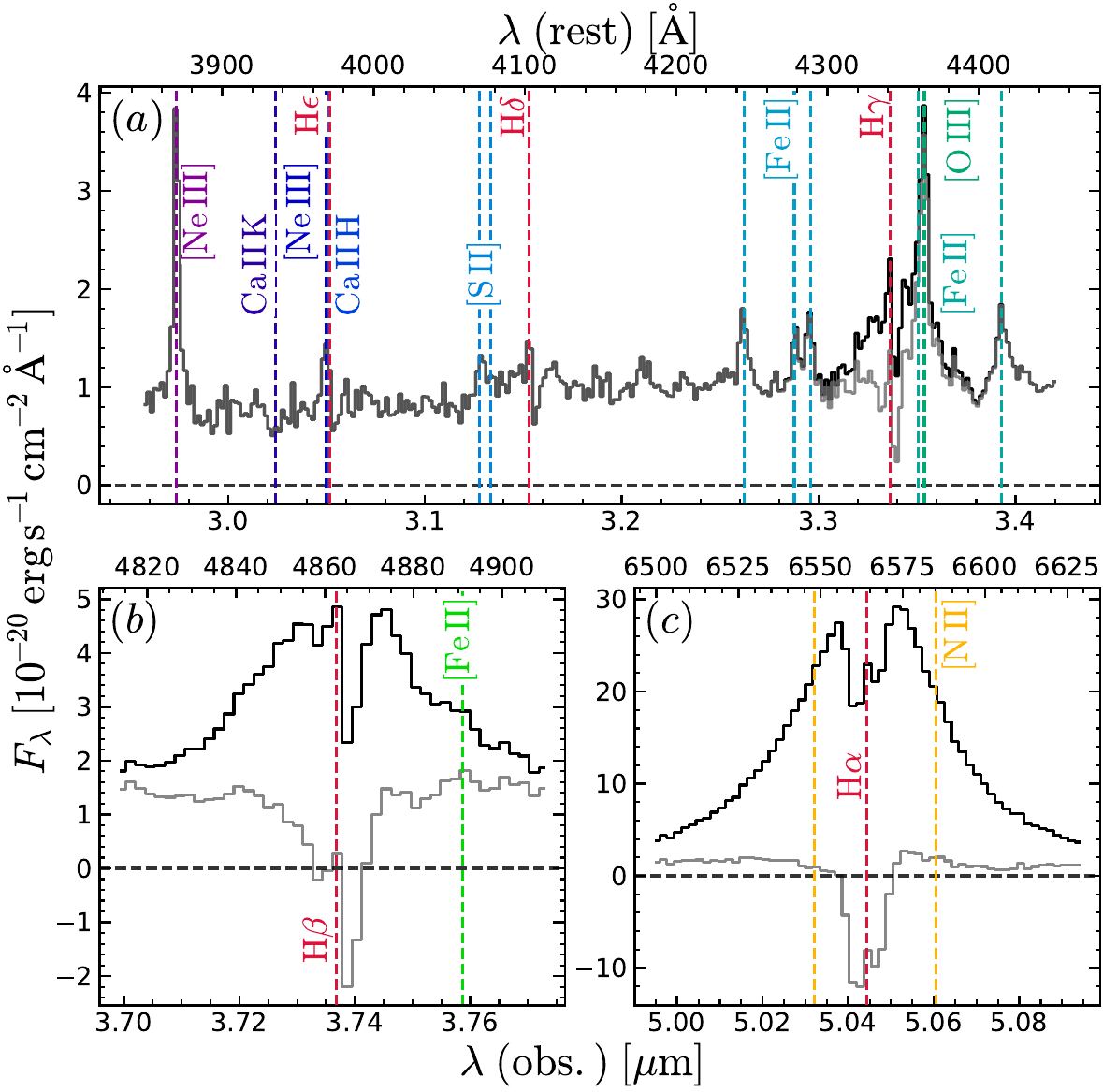}
    \caption{Detail of the \CaII and Balmer absorption features. The wavelength of all lines
    is derived from the systemic redshift, which is
    obtained from the narrow lines. The gray line shows the data after subtracting the best-fit
    BLR model (Section~\ref{s.elmodel.ss.broad});
    both the \Hbeta and \Halpha absorptions reach negative, unphysical flux, meaning that the
    absorbers must absorb the BLR light too and
    ruling out a stellar-atmosphere origin.
    \CaIIL matches well the systemic redshift, while \CaIIL[3968] overlaps both \Hepsilon and
    \NeIIIL[3968], making it impossible to
    measure this line. All Balmer absorption lines
    display a shift; since \Hepsilon--\Hbeta are
    redshifted, while \Halpha is blueshifted, we can
    rule out a single gas screen.}\label{f.absorbers}
\end{figure}

The Balmer-line absorption in \target\ appears to have a non-stellar origin \citep[as seen previously in other systems where the absorption is very near rest-frame velocity][] {deugenio+2025d,deugenio+2025e}.
The \Hbeta absorption reaches deeper than the continuum flux (Fig.~\ref{f.absorbers}), meaning  that subtracting the broad \Hbeta line, would result in an 
unphysical negative flux. For this reason, the absorption cannot be in the continuum only, but must also absorb the broad line emission. This in turn also implies that the absorption cannot arise in the atmospheres of evolved stars.
Therefore, the
absorbing gas must be located between us and the BLR. However, at the same time, the \Hdelta
and \Hgamma absorption are deeper than the underlying broad lines. This implies that
the dense gas is also absorbing the continuum, which must therefore be located behind the
dense gas. This is a natural prediction of models where the continuum is due to direct or re-processed light from an accretion disk \citep[e.g.,][]{ji+2025a}.

We also find evidence that the absorbing gas cannot be arranged in
a passive open screen, with an open geometry, located between the observer and the line-emitting region
Indeed, if that were the case, the optical depths at line center must obey the ratios from quantum mechanics
\begin{equation}\label{eq.oscill}
\begin{array}{cc}
  \tau(\Hepsilon) = 0.012 \cdot \tau(\Halpha) & \;\tau(\Hdelta) = 0.022 \cdot \tau(\Halpha)\phantom{.} \\
  \tau(\Hgamma) = 0.046 \cdot \tau(\Halpha) &  \;\tau(\Hbeta) = 0.138 \cdot \tau(\Halpha).
\end{array}
\end{equation}
In addition, the absorber would likely have a clear minimum, consistent with a
common velocity for all Balmer lines.
But at least two high-redshift LRDs do not satisfy either condition
\citetext{A2744-QSO1, \citealp{ji+2025a,deugenio+2025d}; and
JADES-GS-159717, \citealp{deugenio+2025e}}.
In these past studies, the evidence against the
passive open-screen model was at best marginal, but the exquisite SNR of
\target enables to rule it out completely.
Similar properties have been uncovered in local
LRDs \citep{lin+2025b,ji+2025b}, cementing the hypothesis that these $z=0.1\text{--}0.3$
AGNs are not mere `analogs', but true low-redshift `relics' of the more abundant, high-redshift
population.

In Fig.~\ref{f.absorbers} we compare the location of the absorption dip with the
rest-frame wavelength of each Balmer line; while \Halpha appears clearly
blueshifted, all four lines \Hepsilon--\Hbeta are equally clearly redshifted.
Additionally, the absorption depth increases with decreasing wavelength, opposite
to the predictions of the single passive screen model. A single absorber is thus untenable,
and cannot reproduce the observations.

However, while the double-absorber model of Section~\ref{s.elmodel.ss.broad} can successfully reproduce the
observations, we stress that such model should not be interpreted literally.
Just like we cautioned that the double-Gaussian model for the broad line emission does
not correspond to two physical emitting regions, so the two absorbers do not
imply the existence of two `screens' of gas. Besides, neither of the two absorbers
satisfy the conditions for the relative absorption strengths, Eq.~\ref{eq.oscill},
since we have $\tau_1(\Hgamma) = 1.11\pm0.22$, $\tau_1(\Hbeta) = 1.78\pm0.26$,
$\tau_1(\Halpha) = 2.38\pm0.32$ for the blueshifted absorber, and
$\tau_2(\Hgamma) = 1.48\pm0.30$, $\tau_2(\Hbeta) = 1.47\pm0.22$,
$\tau_2(\Halpha) = 0.13\pm0.05$ for the redshifted absorber.

In principle, the inconsistent velocity and optical depths between the Balmer
absorption lines could be due to faint emission-line components causing
differential line infill. To test this hypothesis, we modified the fiducial
double-Gaussian model to include only a single absorber, but to also have
three separate Balmer-line emitters. These emitters are tied in groups sharing
the same kinematics, $A_V$, and intrinsic ratios from Case-B recombination.
The underlying hypothesis is that the different depth of the Balmer absorption
troughs is multiplicative, so these additive narrow lines can reconcile the line
depths. The outcome however rules out such a model. Statistically, $\Delta\,\BIC
> 500$ relative to the fiducial model. Physically, these emission lines
are relatively narrow ($\sigma = 97\pm12~\kms$ and $107\pm8~\kms$) and bright
($F(\Halpha) = 2\text{--}3.5\fluxcgs[-18]$), yet they have no counterpart in
\OIIIL. Both lines are blueshifted ($-75$ and $-160$~\kms), but the single
absorber is now redshifted ($v_\mathrm{abs}=100\pm10$~\kms).
The narrow profile disfavors an origin in extremely dense regions capable
of suppressing completely \OIIIL.
High attenuation also seems unlikely, since this test model requires $\AV =
11\text{--}12$~mag) -- much larger than any other component of this system.
Hence we reject the hypothesis of a separate emission-line region causing
differential infill.

The different absorption strengths could then be explained by line emission
arising from the same layers where absorption occurs \citep[e.g.,][]{huang+2018,
chang+2025}. However, even in this case, the different redshift suggests that the
absorbing gas must have at least some velocity structure, besides a bulk velocity
relative to the broad lines. An intriguing possibility is that of a turbulent
medium, with inflows and outflows occurring simultaneously, and possibly
related to a `breathing-mode' pulsation of dense gas surrounding the SMBH
\citep[e.g.,][]{deugenio+2025e}.

\subsection{Balmer Decrements }\label{s.elmodel.ss.dust}

The exponential and double-Gaussian models agree on broad-line fluxes, 
yielding Balmer decrements \Halpha/\Hbeta of 9.1--9.2 and \Hgamma/\Hbeta = 0.18. 
By observing three broad Balmer lines, we can infer that no amount of dust 
(assuming standard attenuation laws) can reconcile the line ratios with 
standard Case-B recombination fluxes. The broad lines yield inconsistent 
attenuation levels from \Hgamma/\Hbeta to \Halpha/\Hbeta: standard dust laws 
\citepalias{gordon+2003} would require $\AV = 2.6$~mag from \Halpha/\Hbeta but 
$\AV = 5$~mag from \Hgamma/\Hbeta. Different attenuation laws 
\citep[e.g. the Milky Way's][]{cardelli+1989} can reduce but not eliminate the discrepancy.

In contrast, the narrow lines can be explained by a single attenuation curve 
with modest extinction \citetext{for \citetalias{gordon+2003}, $A_V = 0.47$--$0.48$~mag}. The large broad-line 
decrements are consistent with other LRDs 
\citep{juodzbalis+2024b,wang+2024b,lin+2025b,ji+2025b,deugenio+2025d,deugenio+2025e}, 
but simultaneous three-line detection enables us to definitively rule out 
pure recombination plus dust attenuation, supporting either collisional 
excitation in high-density environments or complex continuum contributions 
consistent with dense gas cocoon models.

\section{Narrow lines}\label{s.narrow}
We detect a wealth of narrow lines in \target including several \FeII lines, as well as auroral \SIIall[4069][4076] and \NIIL[5755] in addition to the usual strong lines. Here we present these detections and their implications for the photoionization source of \target, the physical conditions in its gas, and its dynamical mass. We further discuss detection of \CaII, \NaI, and \FeIIperm  UV absorption. 

\subsection{Modeling approach}\label{s.narrow.ss.modeling}
To model the forbidden lines, we use a spline-based continuum subtraction 
approach that accounts for broad emission-line contamination. We first mask 
the spectrum within $\pm 350~\kms$ from any emission line in 
Table~\ref{t.forb}, then model the remaining spectrum with a cubic spline 
and interpolate over masked regions. This pseudo-continuum includes broad 
emission features (\Hdelta, \Hgamma, \Hbeta) that would otherwise contaminate 
narrow-line measurements.

The emission lines are modeled as Gaussians, and the best-fit model parameters are
estimated using MCMC (Section~\ref{s.elmodel}).
We also include the \NaIall absorber in this model, since its proximity to \HeIL[5785]
makes it necessary to model simultaneously the emission and absorption lines.
For emission-line doublets, we use constrained or fixed flux ratios, as appropriate.
For lines with high critical density (\SIIL[4069], \FeII, \NIIL[5755]) we use a
separate velocity dispersion.

\begin{figure}
\centering
{\phantomsubcaption\label{f.forb.a}
 \phantomsubcaption\label{f.forb.b}
 \phantomsubcaption\label{f.forb.c}
 \phantomsubcaption\label{f.forb.d}}
\includegraphics[width=\linewidth]{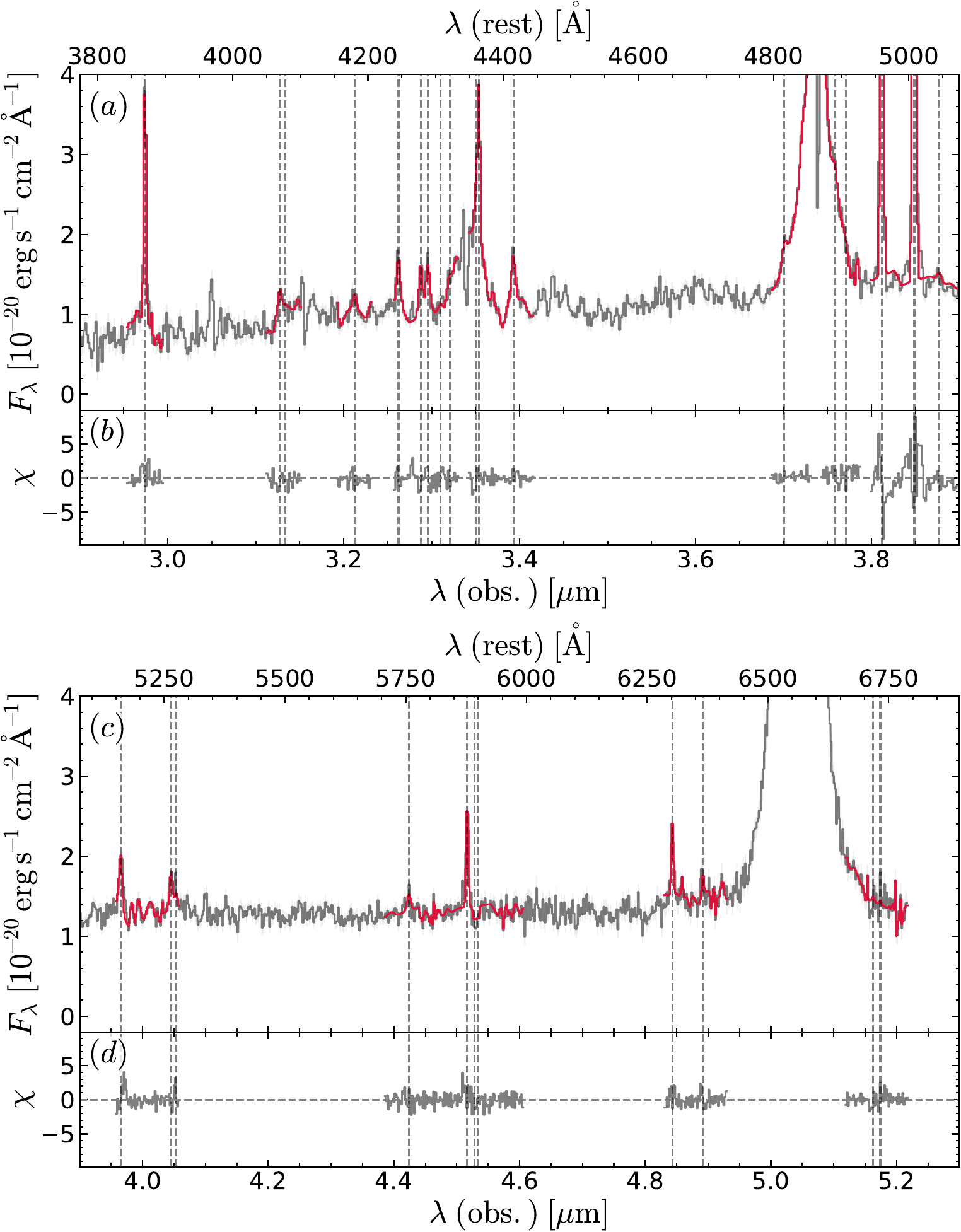}
\caption{Emission-line model for the forbidden lines. The continuum and broad-line
wings are modeled with a spline. We detect several \FeII lines, as well as auroral
\SIIall[4069][4076] and \NIIL[5755]. Auroral \OIIIL[4363] is blended with the fixed-ratio
doublet \FeIIall.
}\label{f.forb}
\end{figure}

\subsection{Line detections and [Fe II] forest}\label{s.narrow.ss.detections}
We detect a wealth of narrow lines in \target, including the first confirmed 
\FeII forest at $z > 6$. The detected \FeII lines include \FeIIL[4245], 
\FeIIL[4277], \FeIIL[4287], \FeIIall, \FeIIL[5159], \FeIIL[5263], and 
\FeIIL[5273]. We also detect auroral lines \SIIall[4069][4076] and \NIIL[5755] 
in addition to standard strong lines (Figure~\ref{f.forb}, Table~\ref{t.forb}).

The fit results are displayed in Fig.~\ref{f.forb} and Table~\ref{t.forb}. Lines with
high critical density prefer a velocity dispersion that 2.5--3 times broader than the
usual strong lines. Some of the reported fluxes are subject to high degeneracies, such
as the nearby lines \FeIIL[4359] and \OIIIL[4363], which also sit on top of broad
\Hgamma, but the overall detection pattern is robust and consistent 
with AGN photoionization in stratified gas environments. 

Recently, \target has also been analyzed by the THRILS team, based on shallower G395M data \citep{lambrides+2025}. 
The two main differences from our conclusions are their reported detection of \FeVIIL, and an extremely high \OIIIL[4363]/\OIIIL ratio. Although
we also detect a strong line at 5,160~\AA, we identify it as \FeIIL[5159] rather than \FeVIIL. We base this
interpretation on three lines of evidence. 
First, we detect several other
\FeII emission lines, whose relative fluxes are consistent with expectation from the Einstein coefficient.
Second, the weakness of metallicity-independent \HeIIL[4686] \citep{wang+2025b} argues against the extreme ionization conditions (IP = 99~eV) required for \FeVIIL emission. If sufficient hard radiation were present to produce \FeVIIL, we would expect brighter \HeII than we observe. Finally, the absence of \FeVIIL[6087], which is typically more luminous than \FeVIIL, is difficult to reconcile with a \FeVIIL interpretation. No other detections of \FeVIIL have
been reported in LRDs, including the independent analysis of a bright LRD from A.~Torralba et al.
(\textit{in prep.}), and even the brightest LRDs at $z=0.1$ \citep{lin+2025b,ji+2025b} and $z=2.26$
\citep{juodzbalis+2024b,ji+2025b}.

As for the very high \OIIIL[4363]/\OIIIL, we interpret the fairly broad narrow line at 4,363~\AA as
a blend of \OIIIL[4363] and \FeIIL[4359] (Table~\ref{t.forb}). In our analysis, \FeII reaches over
80\% of the \OIIIL[4363], hence its contribution cannot be ignored. Simultaneous observation of
other \FeII lines from the same upper level (\FeIIL[4287] and \FeIIL[4414]) enables
inference of the \FeIIL[4359] contribution, breaking any degeneracy with \OIIIL[4363].

A high auroral-to-nebular \OIII ratio has indeed be observed in LRDs \citep{kokorev+2023,jones+2025c},
and is well known in AGN more generally \citep{baskin+laor2005}, where there is independent evidence that
density, not temperature, drives the observed anti-correlation between \OIIIL[4363]/\OIIIL
and \OIIIL/\Hbeta \citep{binette+2024}. However, in these cases (including for UNCOVER~20466),
there is no reported \FeII emission at 4,287 and 4,414~\AA (or any other wavelength), making it both possible and reasonable to attribute the observed line to \OIIIL[4363]. In contrast, the rich \FeII forest we observe in \target necessitates careful accounting for \FeII blending before inferring extreme \OIII ratios.

\subsection{Emission-line ratios and photoionization}

\begin{figure}
\centering
{\phantomsubcaption\label{f.bpt.a}
 \phantomsubcaption\label{f.bpt.b}
 \phantomsubcaption\label{f.bpt.c}
 \phantomsubcaption\label{f.bpt.d}}
\includegraphics[width=\linewidth]{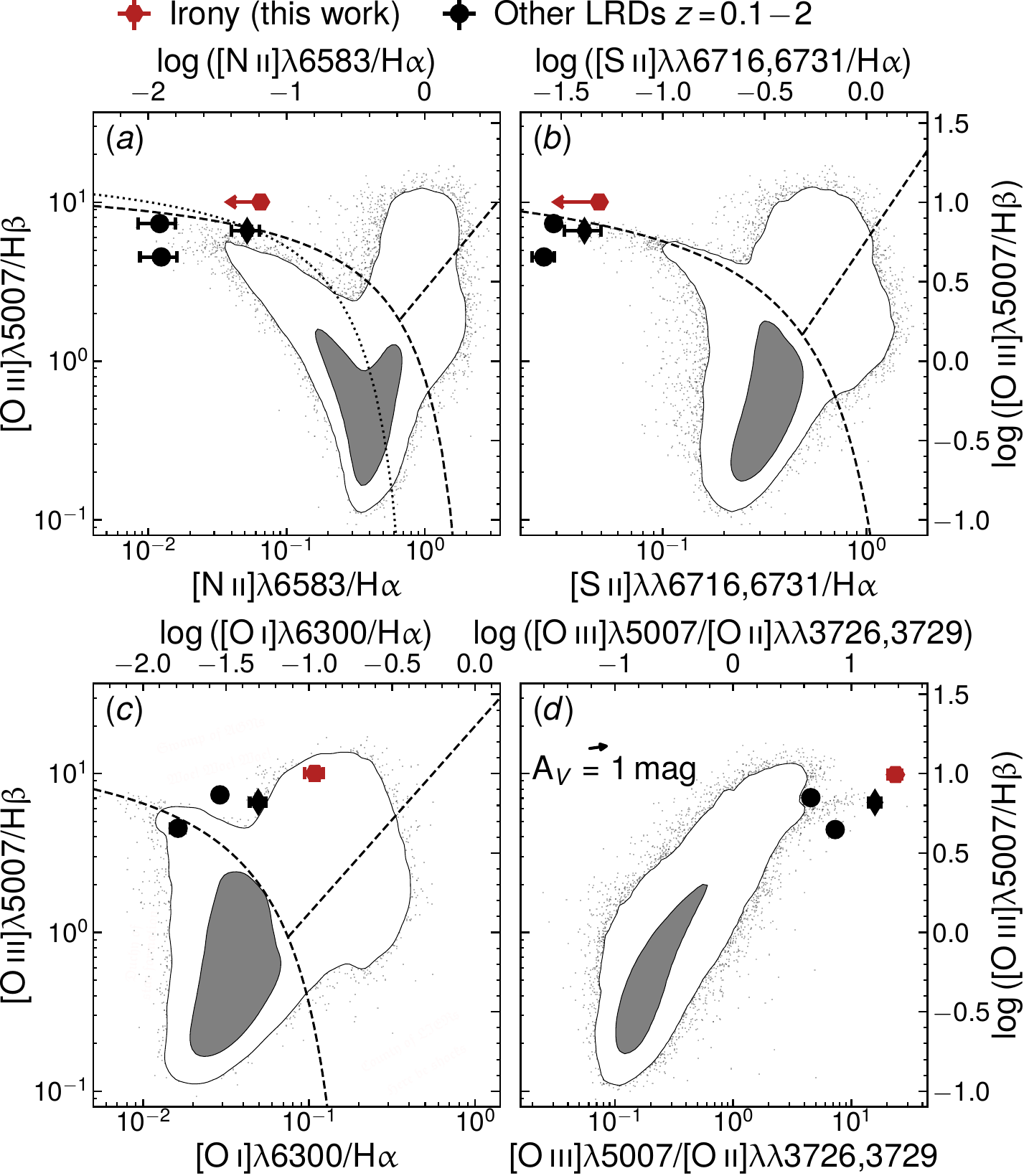}
\caption{Classic emission-line ratio diagrams, including the BPT \citep[panel~\subref{f.bpt.a};][]{baldwin+1981} and VO diagrams \citep[panels~\subref{f.bpt.b} and~\subref{f.bpt.c};][]{veilleux+osterbrok1987}. \target is the red diamond, while the circles are
LRDs from the literature \citep{lin+2025b,ji+2025b,juodzbalis+2024b}. The contours are
SDSS emission-line sources. The \OIL/\Halpha diagram remains surprisingly effective
at singling out AGNs \citep{juodzbalis+2025}; LRDs also seem to occupy an extreme region in the \OIIIL/\Hbeta vs \OIIIL/\OIIall diagram (panel~\subref{f.bpt.d}).}
\label{f.bpt}
\end{figure}

\begin{figure}
\centering
\includegraphics[width=\linewidth]{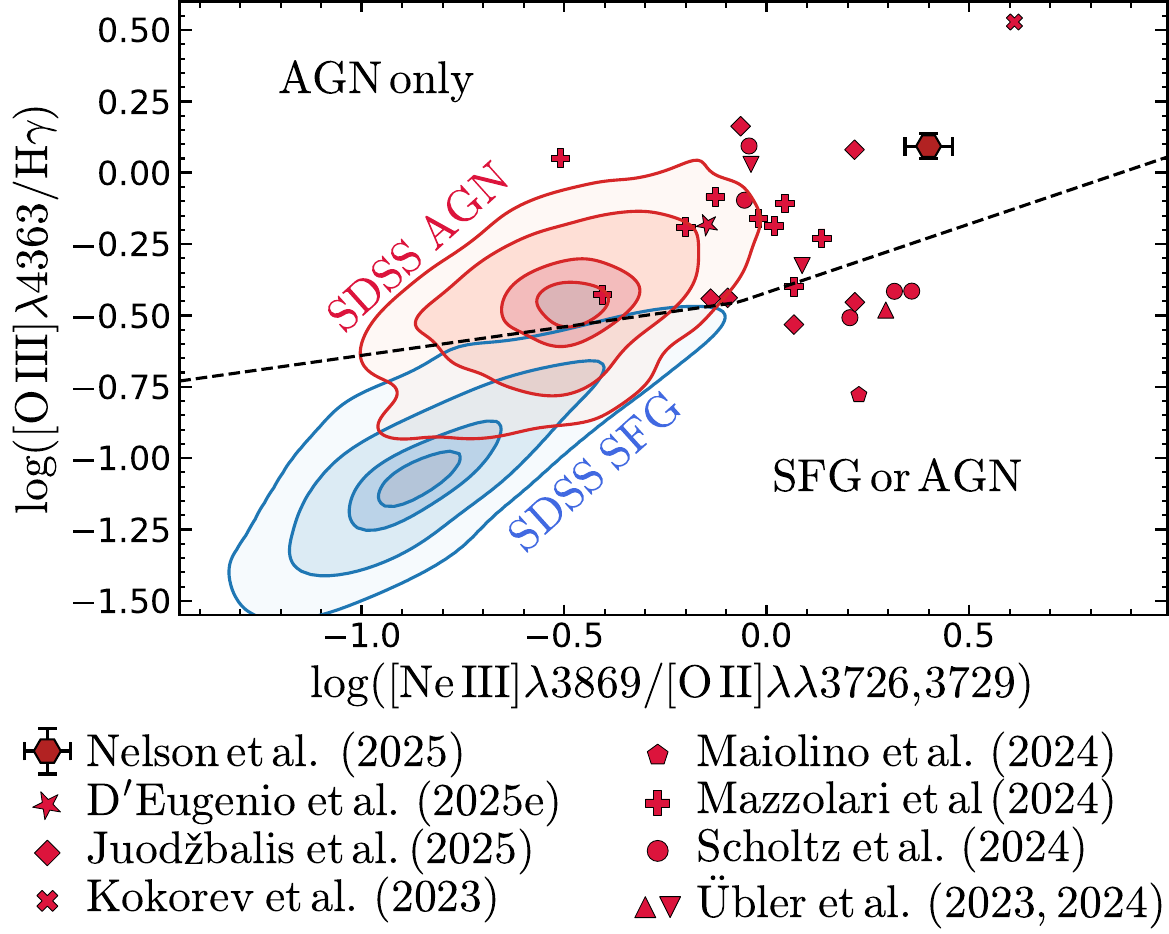}
\caption{AGN diagnostic diagram of \citet{mazzolari+2024}, showing that \target
occupies the AGN-only region, above the dashed demarcation line. For comparison,
we show low-luminosity, high-redshift AGN \citep{maiolino+2024a,ubler+2023,ubler+2024,
scholtz+2025,juodzbalis+2025,kokorev+2023,mazzolari+2024,deugenio+2025e}, while
the contours trace the distribution of star-forming galaxies and AGN from SDSS.
}\label{f.mazzolari}
\end{figure}

The list of emission-line fluxes in Table~\ref{t.forb} can be used to diagnose the photoionization
source of \target. In Fig.~\ref{f.bpt} we present four classic diagnostics diagrams, including
the BPT diagram \citetext{\citealp{baldwin+1981}; panel~\subref{f.bpt.a}},
the VO diagrams \citetext{\citealp{veilleux+osterbrok1987}; panels~\subref{f.bpt.b}
and~\subref{f.bpt.c}}, and the \OIIIL/\Hbeta vs \OIIIL/\OIIall diagram (panel~\subref{f.bpt.d}).
In agreement with the literature, the BPT diagram does not conclusively identify
\target as an AGN \citep[e.g.][]{kocevski+2023,ubler+2023,harikane+2023}. On the other hand, the VO diagram
with \OIL  \citep{juodzbalis+2025} places \target solidly in the AGN region.
In the final diagram (Fig.~\ref{f.bpt.d}), this and other LRDs seem occupy a region that is poorly populated by local
galaxies and AGN (SDSS contours), but is consistent with low-metallicity star-forming
galaxies at $z\gtrsim5$ \citep{cameron+2023}. Fig.~\ref{f.mazzolari} shows the photo-ionization diagram using
\OIIIL[4363]/\Hgamma \citep{mazzolari+2024}, a powerful, dust-insensitive
tracer for high-redshift AGN.  This diagnostic diagram strongly indicates an AGN origin of the narrow lines
in our source, which is confidently placed in the `AGN-only' region.

\subsection{Physical properties of the gas}
The rich array of detected emission lines in \target enables us to probe the physical conditions of the ionized gas through multiple density- and temperature-sensitive line ratios, revealing a complex, stratified structure.
Our measurements and derived constraints are shown in a multitude of colors and line styles in Fig.~\ref{f.temden}.

Different density-sensitive line ratios imply strikingly different electron densities. 
The \OIIall nebular ratio implies a density $\nelec=420~\percm$ (Fig.~\ref{f.temden};
red line). 
However, auroral-to-nebular ratios paint a dramatically different picture.
Although \SIIall remains undetected, the auroral ratio
\SIIall[4069][4076]/\SIIall is also a density tracer,particularly above $\nelec
\gtrsim 10^4~\percm$. We find $F(\SIIall[4069][4071])/F(\SIIall)>1.76$ (3~\textsigma;
Fig.~\ref{f.temden}, dashed black line), implying 
$\nelec>4\times10^4~\percm$, at the 3-\textsigma level (assuming $\Telec=10,000\text{--}20,000$~K) -- two orders of magnitude higher than the value inferred
from \OIIall. 

We also detect the auroral line \NIIL[5755]
\citep[detected independently by][]{tang+2025a}, while the nebular
counterpart remains undetected, which results in an even larger inferred density. The observed ratio $F(\NIIL[5755])/F(\NIIL)>0.5$
requires $\nelec \gtrsim 2.5\times10^5~\percm$ (assuming extremely high
$\Telec\gtrsim30,000$~K; Fig.~\ref{f.temden}, solid green line) -- nearly three orders of magnitude higher than the \OIIall estimate. 

From the auroral-to-nebular \OIII ratio, we also infer a high gas density
(Fig.~\ref{f.temden}; solid blue line). This is because the
flux ratio is $F(\OIIIL[4363])/F(\OIIIL)=0.057\pm0.003$, which meets the \OIIall
locus at a very high temperature $\Telec \approx 24,500$~K. Instead, for the most
stringent lower limit on \nelec (from the nitrogen ratio), we infer a much more
reasonable temperature of $\Telec \lesssim 14,100$~K (where the solid blue and green
lines meet, i.e. at $\nelec \gtrsim 6.3\times10^5~\percm$). 
Of course, N$^+$ and
O$^{++}$ have different ionization potentials, so there is no strong reason for
associating \nelec from N$^+$ to \OIII emission, and thus to prefer
$\Telec \lesssim 14,100$ to $24,500$~K. But the converse is also true, since O$^+$ and
O$^{++}$ need not be cospatial any more than N$^+$ and O$^{++}$. 

The high \nelec implied by the \NII and \OIII line ratios is consistent with the
non-detection of both \SIIall and \NIIall, since both these doublets have lower
critical densities $n_\mathrm{crit}$ (vertical lines below Fig.~\ref{f.temden}). In
contrast, all the auroral lines \SIIall[4069][4076], \NIIL[5755] and \OIIIL[4363] have
higher $n_\mathrm{crit}$ (values higher than $n_\mathrm{crit}>10^7~\percm$ are out of
scale in the figure). (Note that dust reddening has minimal impact on our conclusions,
as indicated by thin dotted lines in the figure.)
The high \nelec we infer is much higher than the value
from the \OIIall doublet; moreover, \nelec is also orders of magnitude higher than
$n_\mathrm{crit}$,
which should suppress \OIIall emission altogether.

The simplest explanation for this combination of observations is a
stratified gas structure, with both low- and high-density regions. 
In addition to the pattern of emission line detections and non-detections, a stratified gas structure is also supported by kinematic evidence.  
The \FeII lines and some of the auroral lines have line profiles that appear distinctively
broader than other forbidden lines ($\sigma = 170\pm10~\kms$ vs $57\pm1~\kms$;
Table~\ref{t.forb}). Since both \FeII and auroral lines have high $n_\mathrm{crit}$,
this kinematic similarity suggests the presence of an inner region compared to the standard
narrow-line region; this inner region is characterized by higher dispersion, high electron
density, and -- surprisingly -- by low-ionization lines. The latter finding is in agreement
with similar findings from local LRDs \citep{lin+2025b,ji+2025b}.

\begin{figure}
\centering
\includegraphics[width=\linewidth]{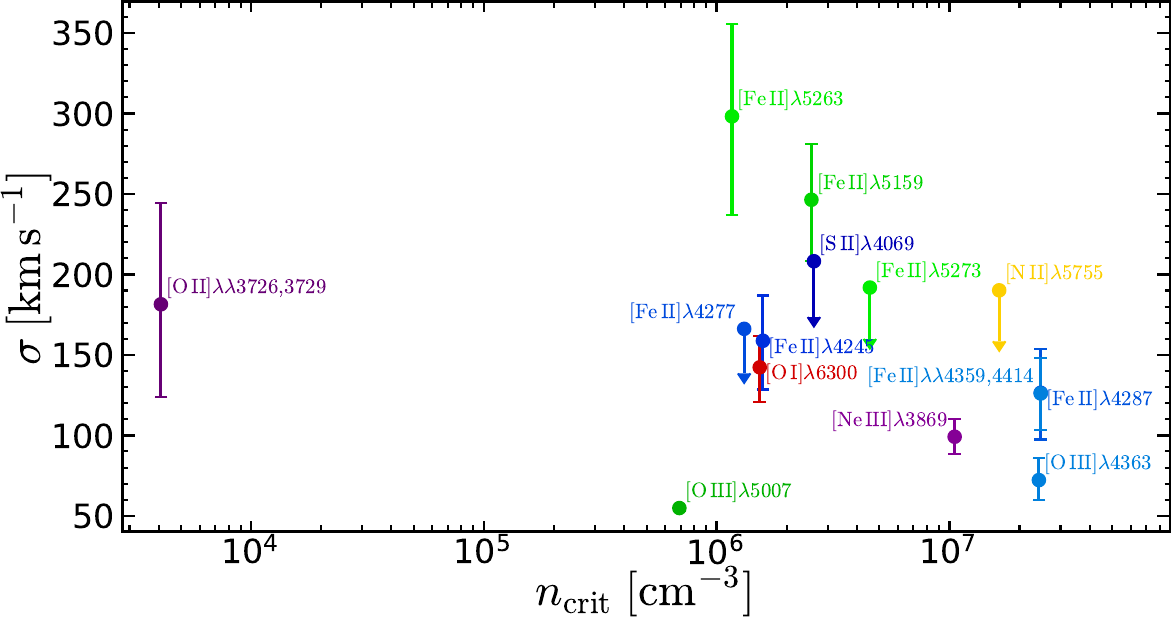}
\caption{We find no clear relation between the line widths and their critical density.}\label{f.critdens}
\end{figure}

\begin{figure}
\centering
\includegraphics[width=\linewidth]{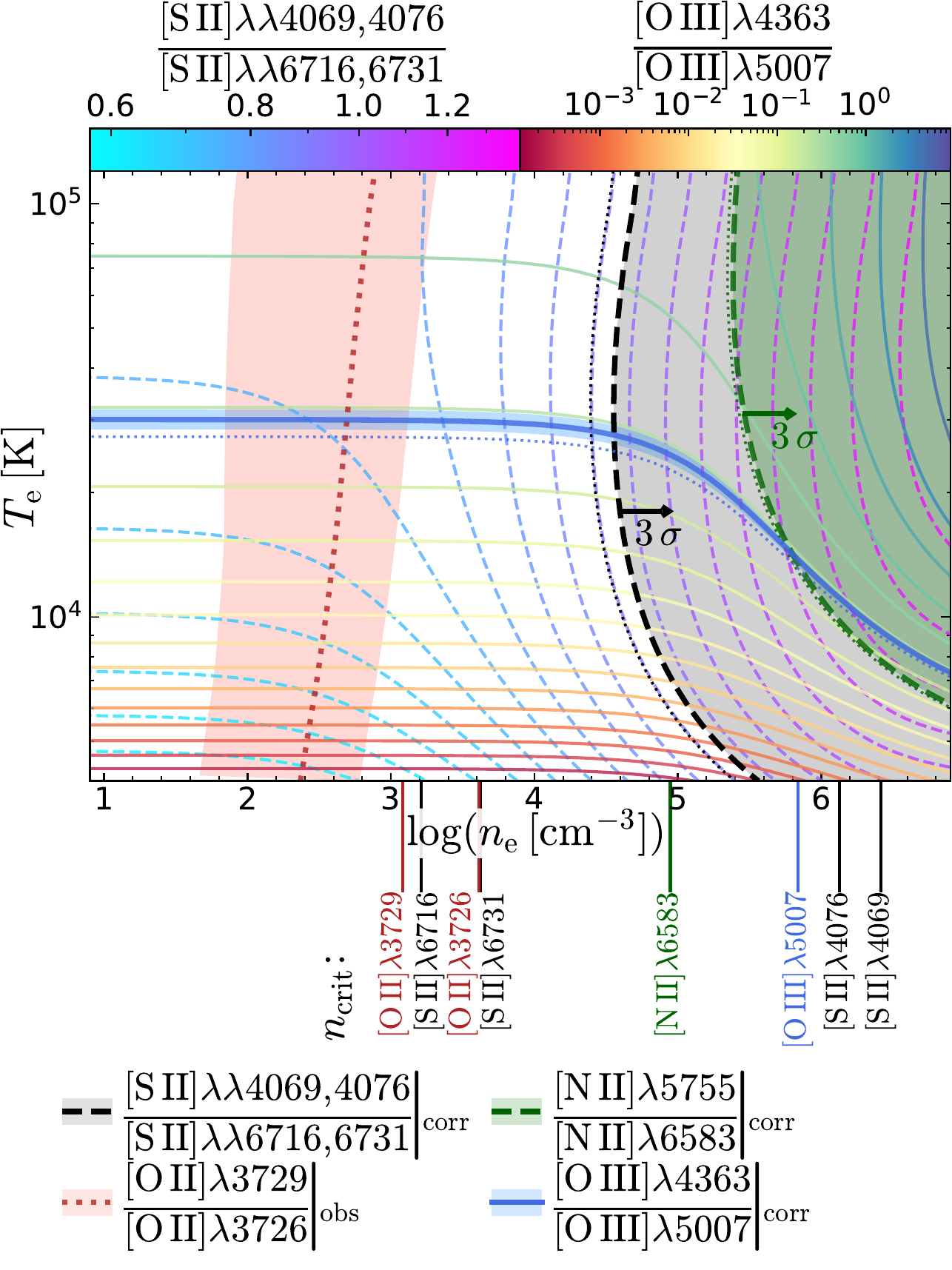}
\caption{The line-emitting gas in \target displays a range of densities, as traced
by the four line ratios considered here (the dashed line is a lower limit, dotted
lines are observed ratios, uncorrected for dust reddening, and solid lines are
reddening-corrected ratios). Critical densities are indicated below (missing lines
have $n_\mathrm{crit}>10^7~\percm$). The \OIII and \NII ratios are best explained by very
high densities, $\nelec > 10^6~\percm$ -- also consistent with the lower limit on
the \SII ratio. These values are much larger than \nelec inferred from \OII, and
larger still than the critical density of either \OII line. This density mismatch
indicates that the gas is stratified, and, therefore, the \OIII lines
must be distributed between the low- and high-density regions.
}\label{f.temden}
\end{figure}

A classic test of stratification is the relation between the width of the lines
(as a probe of the depth of the gravitational well and hence distance from the SMBH), and
their ionization and critical density. In Fig.~\ref{f.critdens} we show this diagram for
\target. To this end, we repeat the narrow-line fit by adding as free parameter the
velocity dispersion of all $SNR>3$ emission lines. When the resulting velocity dispersion
is consistent with 0, we consider the fit outcome as an upper limit, given that the line
was detected with uniform $\sigma = 57~\kms$ or $\sigma = 170~\kms$ in the previous fit
(Table~\ref{t.forb}).

The best detected lines (\NeIIIL, \OIIIL[4363], \OIIIL, \FeIIall and \FeII[4287] form
an increasing trend: lines with higher critical densities are systematically broader, as expected if they arise from progressively deeper in the gravitational potential. Some well detected lines go in the opposite direction. Among these,
\OIIall is not surprising, since this doublet is collisionally suppressed in the
highest-density regions, and must therefore arise from external regions, where other
sources of energy may be at play. The interpretation for \FeIIL[4245] and \OIL is unclear,
and may require better SNR and/or spectral resolution to disentangle.
In particular, high-quality integral-field observations could clarify if any of these
emission lines are spatially resolved.

We note that this stratified picture
complicates our interpretation of the auroral-to-nebular ratios. If a low-density region exists, one must apportion some nebular
emission to it, which would lead to even higher auroral-to-nebular line ratios --
particularly for \OIII, where we cannot invoke chemical stratification \citep[which
instead is possible for \SII and \NII, where nebular emission is undetected;][]{ji+2024}.
Allocating some of the nebular \OIIIL flux to the low-density region would further increase
the auroral-to-nebular \OIII ratio, resulting in an even higher value of \nelec,
comparable but still below $n_\mathrm{crit}$ for several \FeII emission lines.

With the fiducial values of $\Telec=14,100$~K and $\nelec = 6.3\times10^5~\percm$,
we infer a metallicity of $12 + \log(\mathrm{O/H}) = 8.37$, roughly half solar (this neglects
the negligible contribution from $\mathrm{O\,\textsc{ii}}$). Clearly, this is a very
high value, which could vary substantially if the physical conditions \Telec and \nelec
we used are not representative; for instance, replacing the high-density estimate
with $\nelec = 420~\percm$ from \OIIall we would obtain $12 + \log(\mathrm{O/H}) = 8.1$.
Finally, assuming $\Telec = 24,500$~K and $\nelec = 420~\percm$, we would infer
$12 + \log(\mathrm{O/H}) = 7.59$.

In sum, the emerging physical picture of the gas in \target is that of a stratified system -- certainly in
density, perhaps in chemical abundance too, reaching at least
$\nelec \sim 6.3\times10^5~\percm$.

\begin{table}
\centering
  \begin{tabular}{l|cc}
  \hline
    Line                  &   $\sigma$ & $F$           \\
                          &     \kms   & \fluxcgs[-18][] \\
  \hline
    \OIIL[3726]$^\ddag$   & $57\pm1$   & $0.19_{-0.03}^{+0.03}$ \\ 
    \OIIL[3729]$^\ddag$   &    ---     & $0.20_{-0.04}^{+0.04}$ \\
    \NeIIIL               &    ---     & $0.98_{-0.04}^{+0.03}$ \\
    \SIIL[4069]$^\ddag$   & $170\pm10$ & $0.15_{-0.04}^{+0.04}$ \\
    \SIIL[4076]$^\ddag$   &    ---     & $0.04_{-0.01}^{+0.01}$ \\
    \FeIIL[4179]          &    ---     & $0.10_{-0.04}^{+0.04}$ \\
    \FeIIL[4245]          &    ---     & $0.32_{-0.04}^{+0.04}$ \\
    \FeIIL[4277]          &    ---     & $0.29_{-0.03}^{+0.03}$ \\
    \FeIIL[4287]          &    ---     & $0.34_{-0.03}^{+0.03}$ \\
    \FeIIL[4306]          &    ---     & $0.06_{-0.03}^{+0.03}$ \\
    \FeIIL[4320]          &    ---     & $0.02_{-0.02}^{+0.01}$ \\
    \FeIIL[4359]$^\dagger$&    ---     & $0.43_{-0.03}^{+0.03}$ \\
    \OIIIL[4363]          & $57\pm1$   & $0.57_{-0.03}^{+0.03}$ \\
    \FeIIL[4414]$^\dagger$& $170\pm10$ & \FeIIL[4359]/1.436     \\
    \FeIIL[4815]          &    ---     & $0.15_{-0.04}^{+0.03}$ \\
    \FeIIL[4890]          &    ---     & $0.06_{-0.02}^{+0.02}$ \\
    \FeIIL[4905]          &    ---     & $0.01_{-0.01}^{+0.01}$ \\
    \OIIIL[4959]$^\dagger$& $57\pm1$   & \OIIIL/2.98            \\
    \OIIIL$^\dagger$      &    ---     & $11.05_{-0.08}^{+0.09}$\\
    \FeIIL[5044]          & $170\pm10$ & $0.05_{-0.02}^{+0.02}$ \\
    \FeIIL[5159]          &    ---     & $0.45_{-0.04}^{+0.04}$ \\
    \FeIIL[5263]          &    ---     & $0.30_{-0.04}^{+0.04}$ \\
    \FeIIL[5273]          &    ---     & $0.22_{-0.04}^{+0.04}$ \\
    \NIIL[5755]           &    ---     & $0.12_{-0.04}^{+0.04}$ \\
    \HeIL[5875]           &    ---     & $0.52_{-0.03}^{+0.03}$ \\
    \OIL[6300]$^\dagger$  & $57\pm1$   & $0.40_{-0.04}^{+0.04}$ \\
    \OIL[6363]$^\dagger$  &    ---     & \OIL/3.13              \\
    \NIIL[6548]$^\dagger$ &    ---     & \NIIL/3.05             \\
    \NIIL[6583]$^\dagger$ &    ---     & $0.16_{-0.08}^{+0.08}$ \\ 
    \SIIL[6716]$^\ddag$   &    ---     & $0.04_{-0.02}^{+0.02}$ \\
    \SIIL[6731]$^\ddag$   &    ---     & $0.06_{-0.04}^{+0.03}$ \\
  \end{tabular}
  \caption{
  Emission-line properties of \FeII and \SII. We find moderately
  higher $\sigma_\mathrm{n}$ than in Table~\ref{t.broad}, due to
  tying many more lines together in this case.
  $^\ddag$ For doublets arising from the same lower level, we leave
  the flux of the bluest line free, and constrain the flux ratio to
  the range allowed by atomic physics.
  $^\dagger$ For doublets arising from the same upper level, we
  leave the flux of the brightest line free, and use a fixed flux
  ratio to model the faintest line.
    }
  \label{t.forb}
\end{table}


\subsection{Dynamical mass estimate}\label{s.overmassive}

We estimate the dynamical mass of the system from the width of the narrow lines. From the width of the narrow lines, we estimate $\sigma_\mathrm{n} = 55\pm1$~\kms, where
the uncertainties are dominated by systematics in the LSF \citep{degraaff+2024}. We
note that had we used the nominal grating resolution \citep{jakobsen+2022}, the
dispersion would have been even lower ($\approx 45~\kms$). The value is driven
primarily by the strong \OIIIall emission, and remains unchanged when fitting
\OIIIL separately. Simultaneous fit of a broader \OIIIL component is crucial: this
component is statistically required by the data, and ignoring it results in larger
$\sigma_\mathrm{n} = 63~\kms$.

Combining this measurement with the morphology (Section~\ref{s.morph}), we estimate
the galaxy dynamical mass \citep[following the approach outlined in ][]{ubler+2023,
maiolino+2024a}. We adopt the calibration of \citet{vanderwel+2022}, with S\'ersic
index $n=1.8$, $b/a = 0.64$, and $\re = 126$~pc. We correct the observed narrow-line
velocity dispersion $\sigma_\mathrm{n} = 55\pm1~\kms$ upward by 0.175~dex, following
the calibration of \citet{bezanson+2018} to convert gas $\sigma_\mathrm{n}$ to
stellar values. Note that the virial relation of \citet{vanderwel+2022} is calibrated
to reproduce \textit{twice} the mass enclosed inside the sphere of radius \re, following
earlier works \citep{cappellari+2006,cappellari+2013}.
With the numbers above, we obtain $\log(\mdyn/\Msun) = 9.1$. As discussed in \S\ref{s.morph}, the stellar mass for this object based on SED fitting with pre-JWST physical models yields a stellar mass of $\log(\mstar/\Msun) = 11.26$. This high mass estimate is driven by the attribution of the red continuum to starlight. A dynamical mass 2 dex lower than the stellar mass suggests that in this object, the red continuum is likely dominated light from an AGN rather than a mature stellar population (see \ref{sec:cloudy_break} for further discussion).   
Similar to the AGN in \citet{juodzbalis+2025}, also \target lies close to the universal relation between $\sigma_\mathrm{n}$ and \Hbeta luminosity of \HII galaxies \citep{terlevich+melnick1981,chavez+2025}.

\begin{figure}
    {\phantomsubcaption\label{f.overmassive.a}
     \phantomsubcaption\label{f.overmassive.b}}
\includegraphics[width=\linewidth]{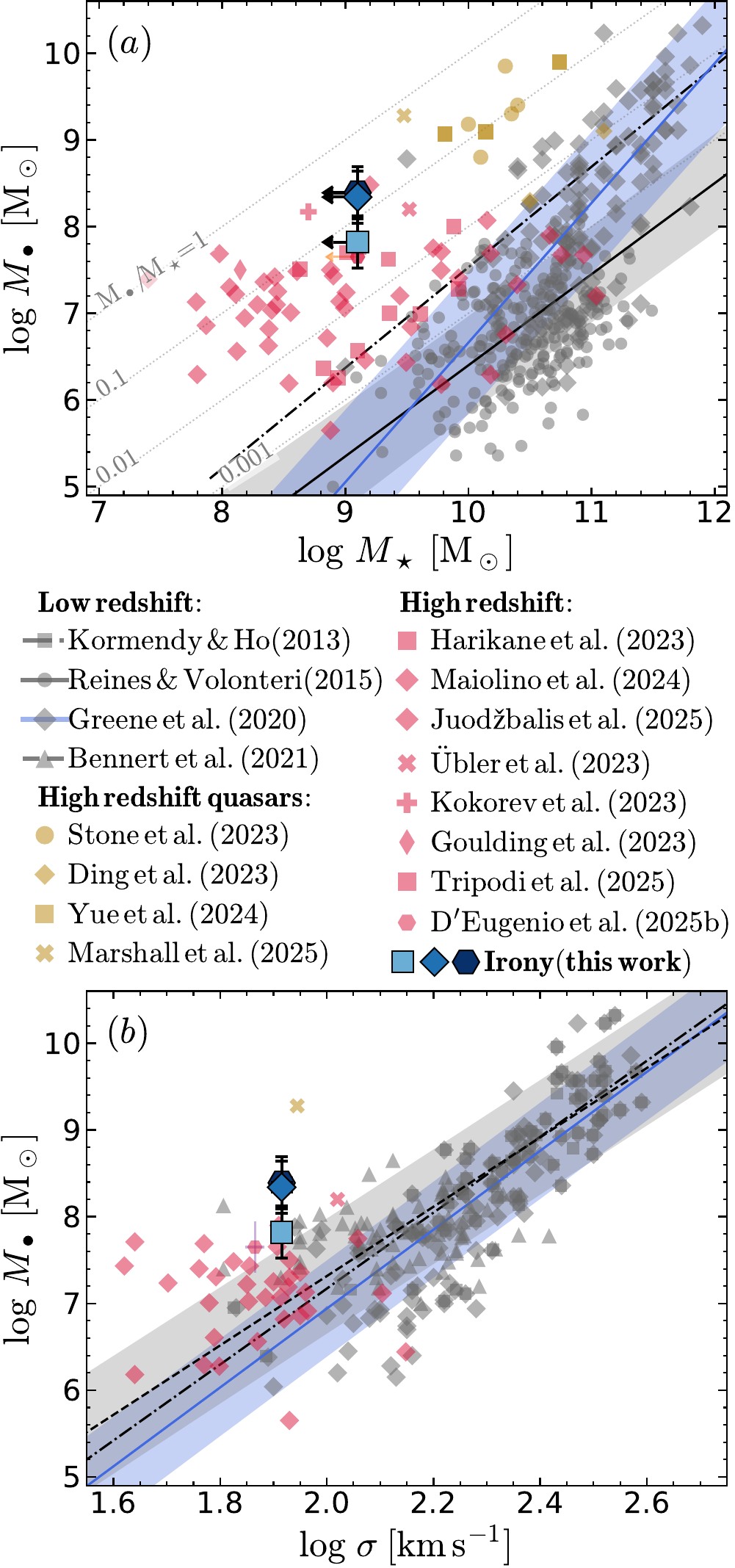}
\caption{We find the SMBH in \target to be overmassive relative to local scaling relations.  high SNR of the \OIIIL emission in \target sets a precise scale on
the host-galaxy velocity dispersion. No matter how we measure the broad-line
width, we obtain SMBH masses that are overmassive compared to local scaling relations.
}\label{f.overmassive}
\end{figure}

\subsection{Absorption lines}\label{s.narrow.ss.abs}

\begin{figure*}
    \centering\includegraphics[width=\textwidth]{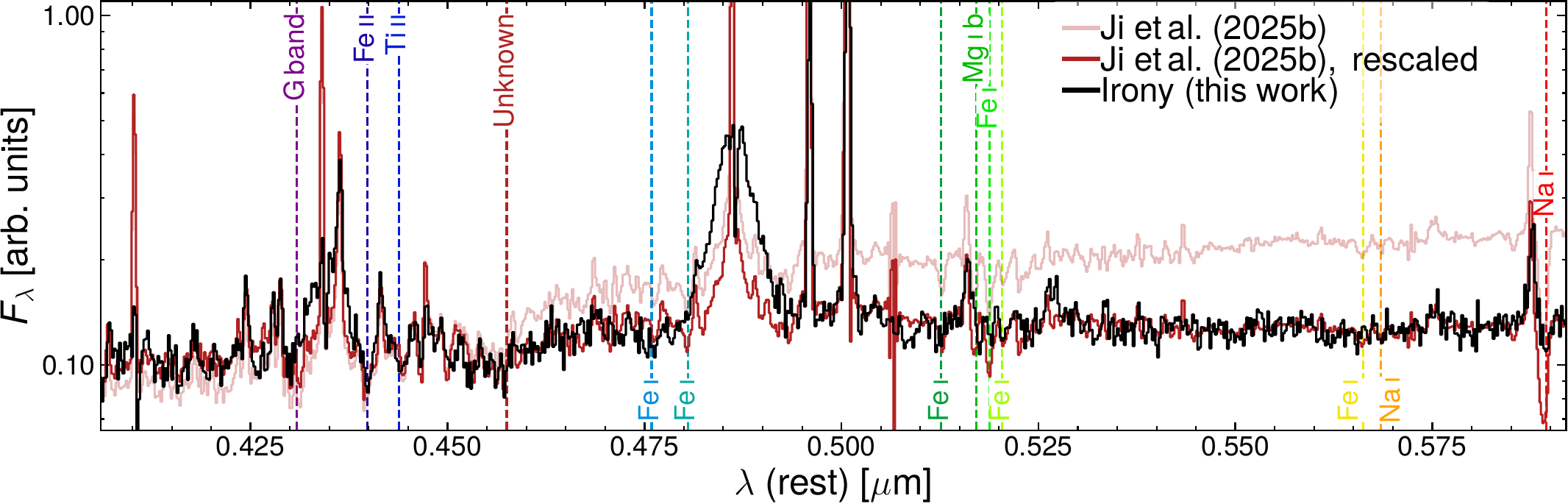}
    \caption{Comparing the continuum of \target (black) and the local `Lord of LRDs' \citetext{red, \citealp{ji+2025b};
    also known as J1025+1407 and \textit{The\,Egg}, \citealp{lin+2025b}}. We find several absorption
    features match between these two objects, pointing to a similar physical origin.
    The GTC spectrum of Lord of LRDs is from zenodo \doi{10.5281/zenodo.17235199}.
    }\label{f.allabs}
\end{figure*}

\target displays a range of absorption features (Fig.~\ref{f.allabs}), many of which match recent
detections at $z=0.1$ \citep{lin+2025b,ji+2025b}. To highlight these features, in Fig.~\ref{f.allabs}
we overplot in light red the GTC spectrum of `Lord of LRDs' \citep[J1025+1407;][]{ji+2025b}; the dark
red curve is spectrum after matching the shape of \target. Some features, such as molecular G-band
and \MgIb, do not find a clear match in \target. Others, such as \FeIIperm, \TiIIperm, and several
\FeIperm lines appear clearly detected. Among the most intriguing features is a distinctive flux
drop near 4,575~\AA, which can be found in many LRDs \citep[e.g.][]{labbe+2024,tripodi+2025,juodzbalis+2024b},
but remains currently unidentified \citep{ji+2025b}.

\begin{figure}
    \centering\includegraphics[width=\columnwidth]{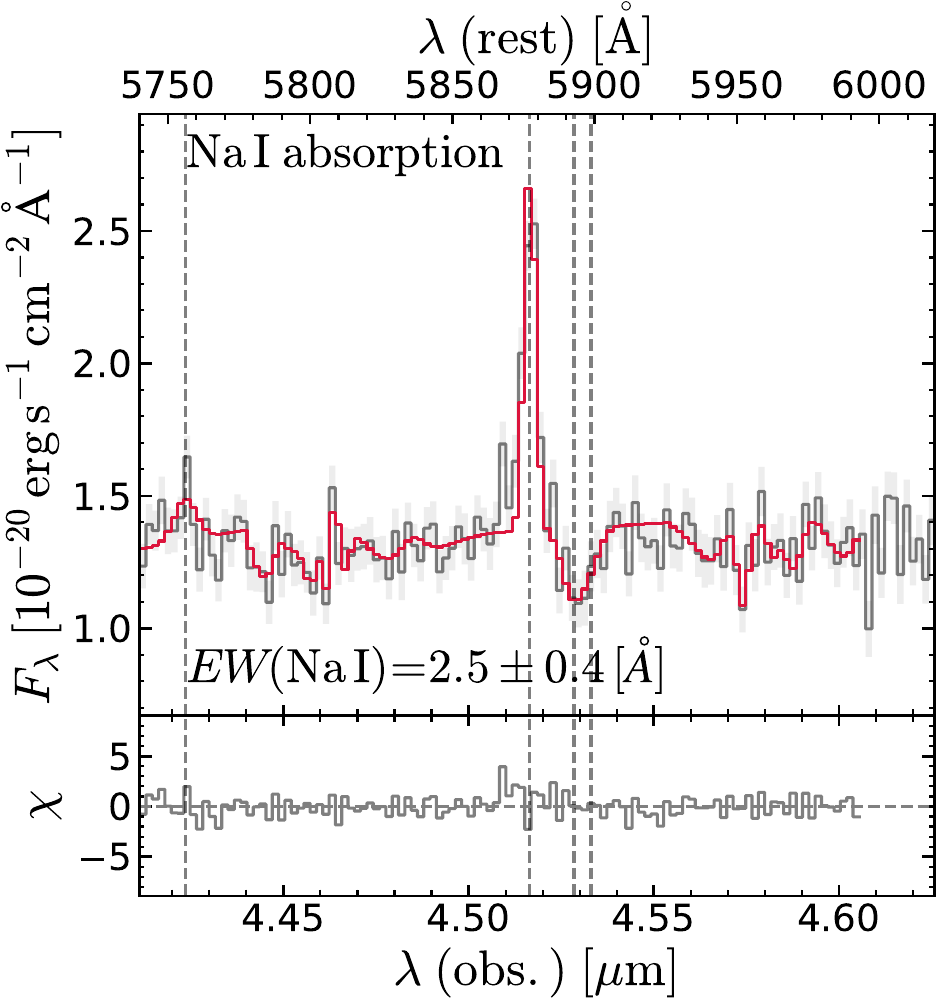}
    \caption{Best-fit continuum and absorption model around the \NaIall doublet.
    The vertical dashed lines mark the rest frame wavelengths of \NIIL[5755],
    \HeIL[5876], and \NaIall.
    We detect clear metal absorption in \target, similar but much smaller
    to what found in local LRDs \citep{lin+2025b,ji+2025b}.}\label{f.abs}
\end{figure}

As part of the narrow-line model, we also model \CaII and \NaI absorption. We use
a standard model with variable velocity, velocity dispersion, covering factor,
and optical depth. Both lines are detected, with $EW(\NaI)=2.5\pm0.4~\AA$ and
$EW(\CaII\,K)=3.0\pm0.7~\AA$. The fit results for \NaI are shown in Fig.~\ref{f.abs}.

\subsection{\FeIIperm UV absorption}

In addition to \CaII and \NaI absorption, we also find \FeIIperm in absorption in the UV. There are three absorption features corresponding to
the well-known \FeIIperm UV1, UV2, and UV3 absorption, previously seen in quasar
absorption spectra \citep{hall+2002} and in star-forming galaxies \citep[e.g.,][]{finley+2017}.
To assess the significance of these features, we model locally the prism spectrum,
fixing the redshift and modeling the \FeIIperm lines as five individual
negative Gaussians, representing \FeIIpermL[2344] (UV3), \FeIIpermL[2374] and
\FeIIpermL[2382] (UV2b and a), and \FeIIpermL[2586] and \FeIIpermL[2600] (UV1b and a).
Given the low resolution of the prism, we do not constrain the optical depth ratios,
nor we attempt to model fluorescent re-emission. Individual lines are not significantly detected, though UV1a approaches significance($EW=14\pm5~\AA$), consistent with the fact that it is expected to be the strongest transition. However, this
does not take into account degeneracies in the model. The posterior probability for
the joint EW of UV1 in $EW(\mathrm{UV1})=22\pm4~\AA$, while $EW(\mathrm{UV2\text{--}3})=23.8\pm5.5~\AA$, supporting a detection. The genuine nature of these lines is also supported by their
independent observation in other high-redshift LRDs: RUBIES-55604 at $z=6.99$ (Fig.~\ref{fig:cloudy_break}, cyan)
and GOODS-N-9771 at $z=5.538$ (A.~Torralba, \textit{in~prep.}). Overall, these may
well be the equivalent of `FeLoBAL' features for LRDs \citep{leighly+2025}.

\FeIIperm UV absorption can arise from the ISM in star-forming galaxies, and from
outflows in quasars. The high EW values we report are intermediate between the two
regimes, since star-forming galaxies can reach $EW\approx 1\text{--}3~\AA$
\citep{rubin+2009,finley+2017}, while quasars can reach 10's of \AA
\citep{rafiee+2016,rodriguez-hidalgo+2011}.
At face value, the 4-\textsigma detection of such strong absorption might indicate
an AGN origin for the UV continuum. It would also open the exciting possibility
to study feedback in LRDs.
To further understand the physical origin of the \FeIIperm UV absorption, we need
higher-resolution observations capable of reliably measuring their individual EWs
and kinematics.
It is plausible that the UV absorption lines, together with the Balmer break and
the optical absorption lines of \target originate in a complex, cocoon-like
gaseous environment, as recently suggested for LRDs
\citep{lin+2025b,ji+2025b,naidu+2025,rusakov+2025}.
As we will describe in the next section, our \cloudy fiducial model produces two strong absorption features in the UV that match the observed spectrum.

\section{{Cloudy Modeling}}\label{sec:cloudy_break}

\begin{figure*}
    \centering\includegraphics[width=0.92\textwidth]{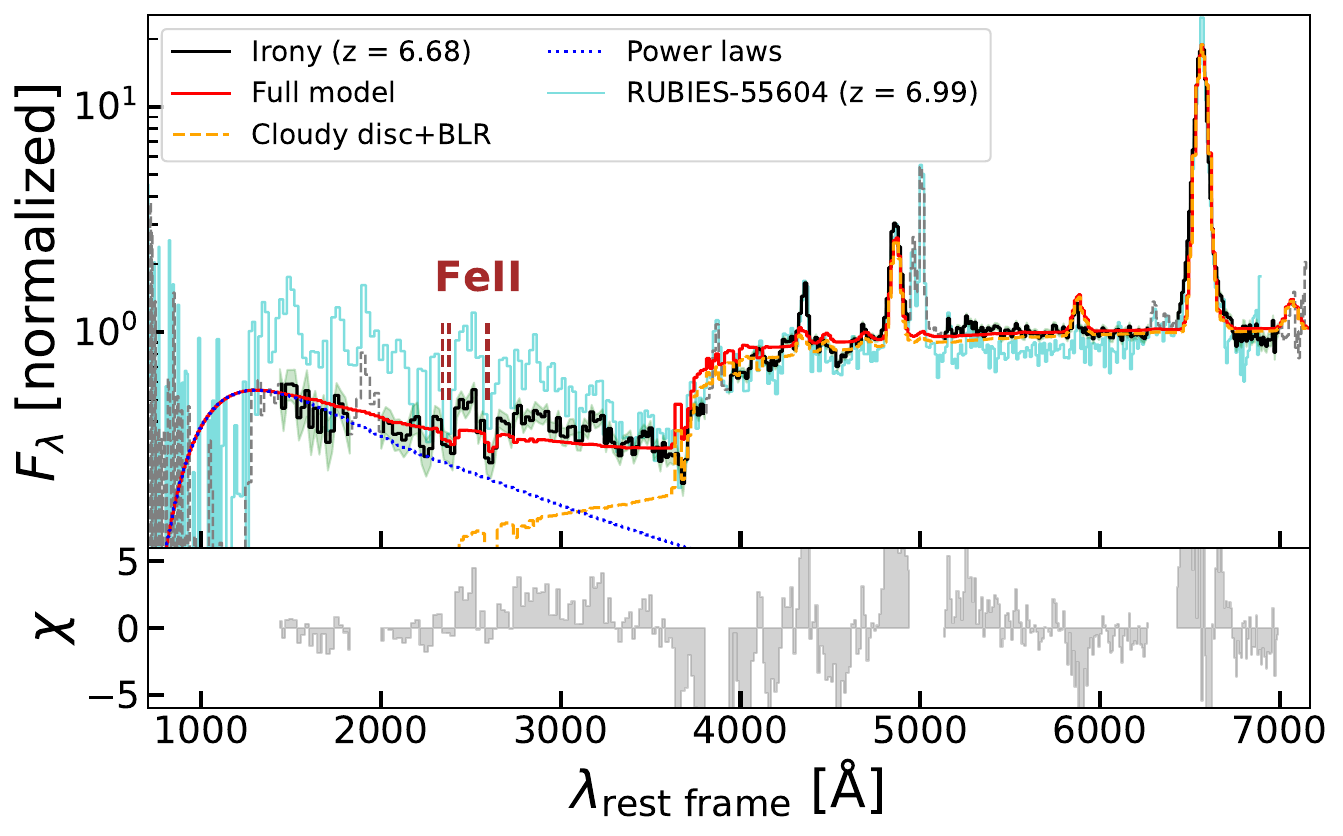}
    \caption{Best-fit continuum + BLR model for \target assuming its intrinsic spectrum is absorbed by dense gas, which creates a Balmer break and Balmer absorption.
    The dashed gray part of the spectrum includes narrow lines that are not fitted.
    The shaded green region on the top panel indicates $1\sigma$ flux density uncertainties.
    The bottom panel shows the fitting residuals normalized by $1\sigma$ uncertainties.
    The model reproduces the overall V-shaped SED but does not well-describe the UV nor the broad line complexity ($\chi ^2_{\nu}=4.2$).
    Marked absorption features correspond to Fe\,{\sc ii} UV1, UV2, and UV3.
    These Fe\,{\sc ii} absorption features have been observed in quasar absorption spectra, and \target is the first LRD to reveal them. Similar findings have been reported in a paper by A.~Torralba (\textit{in~prep.}), and are seen in individual
    high-redshift LRDs \citep[in cyan, RUBIES-55604;][]{wang+2024b}
    }

    \label{fig:cloudy_break}
\end{figure*}
The peculiar V-shaped continua of LRDs likely result from dense gas envelopes surrounding the accreting black hole, which also produce the observed Balmer absorption \citep{degraaff+2025,naidu+2025,ji+2025a,liu+2025,begelman+dexter2025}. We model the continuum and broad-line emission of \target to test this scenario and constrain the physical properties of the absorbing gas.

We follow the approach of \citet{ji+2025a}, using \cloudy \citep{cloudy17} to generate the absorbed spectrum of the continuum as well as nebular emission from the BLR.
We adopt a \citet{pezzulli2017} SED with $M_{\rm BH}=10^8M_\odot$ and $\lambda_{\rm Edd}=1$  and setting $n_{\rm H}=10^{10}~{\rm cm^{-3}}$, $N_{\rm H}=10^{24}~{\rm cm^{-2}}$, $\log U=-2$, and a micro-turbulence velocity of $v_{\rm turb}=\sqrt{2}\sigma _{\rm abs}=113$ \kms (based on our measurements), and assume plane-parallel geometry. We fit the prism spectrum covering the Balmer break and including the UV. The UV is modeled as a superposition of two power laws with dust attenuation independent from the optical component. For simplicity we mask narrow-line dominated regions in the fit and assumed the broad lines have an intrinsic virial broadening of 3000 \kms.
We vary the covering fraction ($C_f$), the attenuation of the BLR ($A_{\rm V}$), the power-law indices, and normalizations of power laws and \cloudy models. 
Posteriors are estimated using \emcee \citep{foreman-mackey+2013} with 15,000 steps.
We note that, while the intrinsic luminosity of the accretion disc models of \citet{pezzulli2017} have been set by the BH mass and $\lambda _{\rm Edd}$, we did not use the tabulated luminosities as the input parameter, but rather used a free normalization since the BH parameters could deviate from the discrete values sampled by the model SEDs.
Our best-fit normalization factor is $0.77\pm 0.01$, which could be interpreted, to first order, as that $\lambda _{\rm Edd}$ is 23\% lower than what is assumed (i.e., 1).
Taking $L_{\rm bol}=0.77\times L_{\rm bol,~model}$, we calculated the inner radius as $R_{\rm in}=\sqrt{Q_{0}/(4\pi c n_{\rm H}U)}=\sqrt{f_{\rm ion}L_{\rm bol}/(4\pi c n_{\rm H} \langle h\nu \rangle_{\rm ion}U)}\approx 0.6$ pc, which is roughly $1.9\times10^4$ times larger than the nominal thickness of the cloud given by $N_{\rm H}/n_{\rm H}$.
This means our plane-parallel set up is a reasonable approximation.

Fig.~\ref{fig:cloudy_break} shows our best-fit model. 
The overall spectral shape is recovered by the model, with best-fit values for the physical parameters $C_f$ and $A_{\rm V}$ being $C_f=0.77\pm 0.08$ and $A_{\rm V}=1.43\pm 0.01$.
However, the model achieves only $\chi^2_{\nu}=4.2$ due primarily to overly simplistic treatment of the complex broad-line profiles (visible in \Halpha), UV continuum excess and absorption not captured by the smooth power laws, and over-prediction of the Balmer break amplitude as the gas-absorbed AGN component (dashed orange line) is pushed up by the UV power-law component. These limitations suggest the need for more sophisticated modeling of broad-line production mechanisms as well as the geometry and dynamics of the absorbing gas. This will be explored in a future work focusing on photoionization calculations.

\section{Nature of IRONY}\label{s.disc}

Its exceptional brightness notwithstanding, \target appears to be a typical LRD with the canonical spectral and structural features. In imaging, it has a compact size and V-shaped SED: blue in rest-UV and red in rest-optical. In spectroscopy, it has narrow forbidden emission lines and broad permitted lines. Further, the spectrum also displays a number of features that have begun to be uncovered in deep, high resolution spectroscopy: narrow Balmer absorption and a Balmer break that is very strong and uncomfortably smooth for a stellar population.
Because of the remarkable brightness of this object, spectroscopy of a reasonable
depth yields a very high SNR spectrum that has allowed us to uncover additional
spectral features which may be common in other LRDs if their spectra were deep enough:
broad Balmer line emission best described by a double-Gaussian profile, Balmer absorption
with complex kinematics, a forest of auroral and rest-frame optical \FeII lines, and
UV--optical metal absorption lines. Below, we discuss how these properties inform our
view of \target and other LRDs.

\subsection{Mass budget and origin of the continuum}\label{s.disc.ss.mass}
The mass budget of \target constrains both its physical nature and the origin of its rest-optical continuum -- both subjects of intense debate for LRDs.
The high SNR of \OIIIL[5007] enables a robust measurement of the intrinsic line dispersion, which combined with the compact size implies a dynamical mass of $\log(\mdyn/\Msun) = 9.1$. This is in strong tension with $\log(\mstar/\Msun) >11$ obtained from stellar-dominated SED fitting \citep[see \S\ref{s.morph} and][]{wang+2024b} \citep[although see][for an alternative interpretation]{Baggen2024}. 
A low \mstar is also suggested by independent analyses
of LRD clustering \citep{pizzati+2024,matthee+2024b,lin+2025c,arita+2025,zhuang+2025}.

In addition to the low dynamical mass we infer from our observations and the low halo masses inferred from clustering of LRDs in general, our spectral analysis provides three additional pieces of evidence favoring an AGN origin of the rest-optical continuum over a stellar origin.
First,the Balmer break is too smooth to be explained by stellar populations(Fig.~\ref{f.obs});
due to stellar physics, a strong break should also be sharp \citep[e.g.,][]{carnall+2023a}.
On the other hand, such a smooth break as shown here resembles the smooth shape of high-turbulence
breaks associated with AGN \citep{ji+2025a,naidu+2025}; the large $v_\mathrm{turb}=113~\kms$ is 10$\times$
higher than the largest values reported in stars \citep[e.g.][]{deugenio+2025e}.
Second, the lack of matching Balmer absorption at the systemic redshift (Fig.~\ref{f.absorbers}), which
is always seen in stars, due to their much lower $v_\mathrm{turb}$.
Third, the depth
of the Balmer lines requires absorption not only of the broad lines but also of the continuum
as seen by the depth of the \Hdelta and \Hepsilon lines (Fig.~\ref{f.absorbers}) \citep[as pointed
out by ][]{juodzbalis+2024b,deugenio+2025d,deugenio+2025e}.
Taken together, these spectral features strongly prefer an AGN origin of the rest-optical continuum. 

The absorber geometry provides an additional argument against a stellar-dominated continuum. 
The high covering fractions ($C_f$; Table~\ref{t.broad}) would require absorbers extended enough to cover most of the galaxy.
implying absorber sizes of order
$\re \approx 2 \times 120$~pc.
However, given the dynamical mass of \target, the absorbers' dispersions of 80 and
116~\kms (Table~\ref{t.broad}) can only be sustained at distances
\begin{equation}
  R_\mathrm{abs} \lesssim \re \, \left( \dfrac{\sigma_\mathrm{n}}{\sigma_\mathrm{abs}} \right)^2 \approx 25\text{--}50~\mathrm{pc}.
\end{equation}
While short-lived configurations may be possible, the high incidence of absorbers in LRDs
 \citep[e.g.,][]{matthee+2024,lin+2025a} seems at odds with this interpretation.

We therefore favor a compact continuum source, with a spatial scale comparable to the broad-line region rather than the host galaxy.
This scenario is consistent with models invoking direct or attenuated AGN accretion disk emission \citep{inayoshi+maiolino2025,ji+2025a,naidu+2025}, as well as models where cool gas envelopes thermalize and reprocess accretion power \citep{begelman+2006,begelman+dexter2025,liu+2025,lin+2025b}.

The absence of coronal lines \citep[in contrast with the similar-mass SMBH
in GS-3073;][]{ji+2024} and the weakness of \HeIIL[4686] \citep{wang+2025b} both
support scenarios of high covering of the accretion disk. The reported weakness of \Lyalpha in UNCOVER-45924 \citep{torralba+2025} is also consistent with this picture.
However, \Lyalpha statistics for LRDs are still sparse, and some LRDs do display
lines with high ionization potentials \citep{tang+2025a}, and extended \Lyalpha
haloes \citep{morishita+2025}. Deep rest-UV spectroscopy and larger \Lyalpha surveys are needed to understand the properties of the whole population.

\subsection{Iron abundance}

\FeII emission is usually weak in \HII regions \citep{grandi1975,osterbrock+1992},
while in LRDs it reaches remarkably high values of $0.5\,F(\Hbeta)$. Measuring the
gas-phase abundance is not straightforward, due to the different ionization
potentials of \FeIperm and H.
Following \citet{bautista+pradhan1998}, we leverage the similar critical densities
and excitation temperatures of \FeII and \OI to estimate
\begin{equation}
\dfrac{N(\FeIIperm)}{N(\OIperm)} = \dfrac{F(\FeIIL[5179])}{F(\OIL)} \dfrac{j(\OIL)}{j(\FeIIL[5179])}\approx 2
\end{equation}\label{eq.irono}
where $j$ is the emissivity calculated for $\Telec=14,100$~K and
$\nelec=6.3\times10^5~\percm$. Even assuming no dust depletion, this ratio is very
high \citep[the solar abundance ratio across all ions is $\mathrm{Fe/O} =
0.059$;][]{asplund+2009}. Collisional suppression of \OIL seems unlikely, because
this line has fairly large $n_\mathrm{crit}$ (comparable to several \FeII lines,
Fig.~\ref{f.critdens}), and because we do not detect auroral \OIL[5577].
Permitted \OIresL is outside the spectral range \citep{juodzbalis+2024b,tripodi+2025}.

A possible explanation for this large value is that \FeII may be powered mostly by
fluorescence \citep{baldwin+1996}. A comparison with other low-ionization lines
subject to fluorescence such as \OIresL would be illuminating, but is not possible
with NIRSpec for an object at $z=6.68$.
Of course, our estimate assumes a ionization correction factor (ICF) of $\approx 1$;
this is expected since \FeII should arise primarily from the same regions where O stays
neutral. To estimate the contribution of \FeII from regions where O is ionized, we
can compare to \OII emission. Due to the high densities probed by \FeII, we would expect
substantial \OII from lines with high critical density, i.e. auroral \OIIL[2470] and
transauroral \OIIAuall. Intriguingly,
strong \OIIAuall is indeed observed in lower-redshift LRDs
\citep{juodzbalis+2024b,lin+2025b,ji+2025b}, but this line is outside the NIRSpec
range at $z=6.68$. However, the emission-line group near 2,500~\AA in the prism
spectrum may have a strong contribution from \OIIL[2470]; using a line flux of
$F(\OIIL[2470])=0.9\fluxcgs[-18]$, we would infer a much more reasonable
$N(\FeIIperm)/N(\OIIperm)\approx0.09$ (no dust correction) or $\approx 0.055$
(assuming $\AV=0.47$~mag and the \citetalias{gordon+2003} reddening law). These values
are still very high, because Fe is much more refractory than oxygen: typical
depletion of iron can be factors of 1,000, while SNe enrichment is only 10,
hence bright \FeII must be associated with low depletion on dust grains more
than chemical enrichment \citep{spitzer+jenkins1975,phillips+1982}.
Nevertheless, even assuming zero depletion, a near solar Fe/O abundance ratio
seems implausible, because high Fe/O values are associated primarily with long
star-formation histories and enrichment from type-Ia supernovae
\citep{matteucci+greggio1986,maiolino+mannucci2019}, both of which can be ruled
out at $z=6.68$ due to the short age of the universe.

In galaxies with strong forbidden \FeII emission, such as NGC4151, this is attributed to shocks. Shocks
can both destroy dust, and excite \FeII \citep{knop+1996}, as indeed seen in supernovae. Alternative
scenarios, such as photo-evaporation due to the AGN continuum could imply an origin near the BLR, but
the width of the forbidden lines we detect is too narrow to be associated with the BLR itself.
Equally puzzling is the lack of permitted \FeIIperm emission. While this has been reported for low-
luminosity type-1 AGN \citep{trefoloni+2025}, the possibility of low metallicity seems ruled out by our
finding of widespread emission due to forbidden \FeIIperm.

We can draw an intriguing parallel with bright quasars, where past works also found surprisingly
high Fe/O abundance ratios \citep{dietrich2003,maiolino+2003}, which may indicate
rapid enrichment in the nuclear region. Future works analysis the full suite of lines from
\OIresL[1304], \FeII, and \OIresL \citep[e.g.][]{tripodi+2025} would be able to shed light on this issue.

Regardless of the precise Fe/O abundance, the mere detection of \FeII at $z=6.68$ implies
that this kind of emission may be actually widespread in LRDs across redshifts, with reported detections
at $z=0.1$ \citep{lin+2025b,ji+2025b} and $z=2.26$ \citep{ji+2025b}. In hindsight, other bright LRDs with
low-resolution spectroscopy may also display blended \FeII emission
\citep{labbe+2024,tripodi+2025,taylor+2025b}, and definitive evidence will certainly
come from future higher-resolution \jwst programs.
Interestingly, \FeII emission seems more essential than a Balmer break, for
J1025+1402 does display Balmer absorption, \FeII emission and exponential broad lines,
but lacks a Balmer break \citep{ji+2025b} -- a feature which is seen in both
\textit{Rosetta Stone} \citep{juodzbalis+2024b} and in \target, as well as in all the
candidate \FeII emitters \citep{labbe+2024,tripodi+2025,taylor+2025b}.

\subsection{Absorption lines and cool gas envelope}

\citet{liu+2025,lin+2025b} proposed the presence of a cool gas envelope around
J1025+1402, with $T\sim5,000$~K. In their model, the continuum is due to thermal
emission from this envelope, whose low temperature may explain the lack of a
Balmer break. In our case, we observe weak absorption in both \CaII and \NaI.
\NaI in particular appears distinctively weaker than in J1025+1402, suggesting
lower metallicity or higher temperature -- capable of suppressing neutral
sodium. If a gas envelope was present, a higher temperature of around
10,000~K would imply a blue SED peak, whereas the red optical slope in \target
favors lower \Telec. Interestingly, the temperature inferred from a simple
electron-scattering scenario is $\Telec \sim 6,600$~K (Table~\ref{t.broad}),
very close to the value proposed for J1025+1402 \citep{lin+2025b}. On the other
hand, such simple scattering is likely too simplistic, because it fails to
reproduce simultaneously multiple hydrogen lines \citep{brazzini+2025b}.
Still, the model is not fully ruled out, because other radiative-transfer
effects could be at play \citep[e.g.,][]{chang+2025}.

\subsection{Black Hole Mass and Scaling Relations}\label{s.disc.ss.mbh}
The interpretation of broad Balmer lines in LRDs has implications for black hole mass estimates and potentially our understanding of early supermassive black hole growth. The complex line profiles recently observed in \target and other LRDs suggest signficant uncertainties in applying standard mass calibrations to the emission lines in these enigmatic objects.
If LRDs are standard AGN, broad Balmer lines should arise from gas clouds in a broad line region (BLR), orbiting the supermassive black hole at high speeds. While real AGN often show complex broad-line structure requiring multiple Gaussian components, the underlying expectation is that Doppler broadening due to virial motions dominates the line profiles \citep[e.g.][]{kollatschny+2013}.

However, as demonstrated in \S\ref{s.elmodel.ss.results}, our Balmer line profiles show exponential wings that are difficult to explain with virial broadening alone, challenging this standard interpretation. 
\citet{rusakov+2025} proposed that the exponential broad line profiles in LRDs could be explained by electron scattering in dense gas \citep[e.g.][]{weymann1970,laor2006,huang+2018} rather than virial broadening.
\citet{begelman+dexter2025} suggest that the origin of these broad lines may be
electron scattering in the atmosphere of a quasi-star instead of virial broadening
in a broad line region. In this model, LRDs are $\sim10^6~\Msun$ black holes
embedded in and accreting from envelopes of gas whose emission is powered by
accretion onto the central BH at dramatically super-Eddington rates.
However, even assuming electron-scattering drives the observed exponential wings, we still detect an unscattered, broad Gaussian component with $FWHM > 1,000$~\kms (Table~\ref{t.broad}).
This component is perhaps difficult to explain in a pure `quasi-star' model, and
seems to require at least a hybrid model, with substantial virial broadening.

Recently \citet{juodzbalis+2025b} used \jwst/NIRSpec spectroastrometry
\citep{gnerucci+2011} 
to estimate the dynamical mass profile of Abell2744-QSO1,
an LRD at $z=7.04$,
which shares several properties with
\target: a strong Balmer break \citep{furtak+2024} of non-stellar origin
\citep{ji+2025a}, narrow \OIIIL and \Halpha \citep{ji+2025a,deugenio+2025d},
and rest-frame Balmer absorption \citep{ji+2025a,deugenio+2025d} with complex
optical depths and kinematics \citep{deugenio+2025d}.
This independent \mbh estimate agrees with the single-epoch virial calibrations,
provided one uses the broadest Gaussian component of \Halpha in the calibration.
Measuring \mbh from the broadest Gaussian in our double-Gaussian fit, we would infer $\log(\mbh/\Msun) = 8.8$, within a factor
of two from the upper bound represented by \mdyn, and with a sphere of influence
of 400~pc. 

Fig.~\ref{f.overmassive.a} shows \target in the context of observed relations between SMBH mass, host stellar mass, and host velocity dispersion. Regardless of the adopted broad line profile, the inferred
\mbh is large and places \target above local \mbh--\mstar relations \citep{
reines+volonteri2015,greene+2020}. If single-epoch virial relations are correct, this and many other early SMBHs are overmassive relative to their host galaxies
\citep[e.g.,][]{ubler+2023,kokorev+2023,harikane+2023,juodzbalis+2024a,juodzbalis+2025b,maiolino+2024a,marshall+2025,tripodi+2024}. With the large inferred \mbh,
\target bridges the regimes of low-luminosity AGN \citep{harikane+2023,
maiolino+2024a,juodzbalis+2025,goulding+2023,ji+2025a,deugenio+2025e} and quasars
\citep{stone+2023,ding+2023,yue+2024}.
Furthermore, with a narrow line velocity dispersion of $55\pm1$~\kms, \target is also found
above the \mbh--$\sigma$ relation \citep[Fig.~\ref{f.overmassive.b};][]{
kormendy+ho2013,bennert+2021}.
In agreement with
\citet{wang+2024b}, and similar to other LRDs \citep{ji+2025a,ji+2025b,
deugenio+2025d,deugenio+2025e,marshall+2025}, \mbh is a substantial fraction of the dynamical mass in \target,
ranging between $\mbh/\mdyn = 0.06\text{--}0.20$, depending on the broad-line
model adopted.

Of course, a large \mbh is not without difficulties. To start, single-epoch calibrations
yield universally low Eddington rates, which begs the question of where and when does intense
accretion take place \citep[e.g.,][]{rusakov+2025}.
One hypothesis is that the super-Eddington accretion has a very small duty cycle (1--4\%) so it may
be much more likely to observe these AGN in their long sub-Eddington phases than during the short,
bursty phase.
It has been argued that the high implied luminosity density to
accrue such large masses would violate the Soltan argument \citep[e.g.,][]{greene+2025}.
Larger samples of directly measured \mbh may help address these open issues.

\section{Summary and conclusions}\label{s.conc}

We presented deep \jwst/NIRSpec spectroscopy of the bright Little Red Dot (LRD)
AGN \target at $z=6.68$ \citep{wang+2024b}, enabling the first high-SNR,
medium-resolution dissection of its continuum, broad Balmer lines,
multi-component Balmer absorption, and a rich forest of weak forbidden transitions.

\begin{itemize}

\item The data reveal broad Balmer emission from \Halpha through \Hdelta, and
concurrent $n=2$ hydrogen absorption in \Halpha--\Hepsilon.
The Balmer decrements of the narrow lines are consistent with modest dust attenuation
($\AV=0\text{--}0.5$~mag, model dependent; Table~\ref{t.broad}). In contrast, the broad Balmer decrements cannot be reconciled with standard dust attenuation applied to
Case-B recombination, requiring a substantial collisional component to the excitation in order to explain the high \Halpha/\Hbeta$\sim9$ (Fig.~\ref{f.dustatt}).

\item The classic \OIL/\Halpha emission-line diagnostic places \target firmly in
the AGN regime, as does the \citet{mazzolari+2024} diagram based on \OIIIL[4363].

\item The Balmer absorption is extremely deep
(Fig.~\ref{f.absorbers}). To avoid unphysical negative flux, the depth of
\Halpha and \Hbeta requires absorption of the BLR light; at the same time,
the depth of \Hdelta and \Hepsilon requires absorption of the continuum too.
These facts rule out a stellar origin, and place the absorbers firmly between
the observer and \textit{both} the BLR and the continuum (including at the outer
edge of the BLR).

\item The Balmer absorbers are intrinsically complex (Fig.~\ref{f.absorbers}),
exhibiting kinematics and optical-depth ratios that are inconsistent with each
other. \Halpha has a blueshifted trough, while higher-order lines
(\Hbeta--\Hepsilon) are redshifted. Higher-order lines have larger measured
optical depth than \Halpha, confirming earlier indications \citep{deugenio+2025d,
deugenio+2025e}, that these optical depths do not follow quantum-mechanical
ratios.

\item A passive screen, with an open geometry, located between the observer
and the line-emitting region is therefore untenable. Instead, at least
two absorbing components with different velocities, covering factors, and optical
depths are required, consistent with dense gas clouds situated between us and the
BLR and partially covering both the BLR and the continuum-emitting region.
This geometry independently rules out an origin in evolved stellar atmospheres and
supports scenarios in which dense circumnuclear material shapes both the continuum
and line transfer in LRDs.

\item We compared three BLR profile models: an exponential profile motivated by
electron scattering, a double-Gaussian effective profile, and a Lorentzian
(Voigt) profile. All three models are fitted simultaneously to \Hgamma, \Hbeta
and \Halpha within a Bayesian framework (Fig.~\ref{f.fit}).
While the line wings look strikingly exponential, the overall fit quality is
dominated by subtle asymmetries around \Halpha, which favor the double-Gaussian
model. However, if those asymmetries are suppressed (e.g., by tying the
double-Gaussian centroids), the exponential model becomes preferred. The adopted
profile matters: the inferred FWHM ranges from $1350~\kms$ (exponential) to
$2580~\kms$ (Lorentzian), which propagates to single-epoch virial black-hole
mass estimates that differ by a factor of four ($\log(\mbh/\Msun)=7.82\text{--}8.39$)
and Eddington ratios spanning $\lambda_\mathrm{Edd} = 0.4\text{--}1.7$.

\item The forbidden lines are remarkably narrow ($\sigma = 55~\kms$), implying a small
dynamical mass and, hence, a high \mbh/\mdyn ratio, strengthening the case that
some LRDs host over-massive black holes relative to their hosts, even in the regime
of large \mbh.

\item The forbidden-line spectrum is remarkably rich in low-ionization,
high-critical-density transitions: \SIIall[4069][4076], \NIIL[5755], and numerous
\FeII lines are detected, with the high-$n_\mathrm{crit}$ lines systematically
broader ($\sigma \approx 170~\kms$) than standard narrow lines.
Line-ratio diagnostics that include \OIIIL[4363]/\OIIIL and limits from
\NIIL[5755]/\NIIL imply very high electron densities, $\nelec \gtrsim 6.3\times
10^5~\percm$, while the \OIIall doublet indicates a much lower-density zone
($\nelec \approx 420~\percm$). The inevitable conclusion is a stratified
narrow-line region in density (and possibly chemistry), with a compact, dense inner
zone that powers \FeII, auroral \SII and \NII, superimposed on a lower-density
region producing \OIIall. This structure helps explain both the coexistence of
high- and low-$n_\mathrm{crit}$ lines and the distinct kinematics of the
\FeII group. The prominence of forbidden \FeII at $z = 6.68$, together with weak
\CaII and \NaI absorption, hints at low depletion or shock-related dust processing,
though a detailed Fe/O budget remains uncertain without trans-auroral and UV
constraints.

\item A simple absorbed-AGN continuum model (\cloudy-based) reproduces the
global V-shaped SED and Balmer break with a high covering fraction ($C_f
\approx 0.77$) and moderate attenuation ($A_V \approx 1.4$~mag), while leaving
structured residuals around the broad lines and in the UV. The latter include
\FeIIperm UV1--UV3 absorption complexes, suggestive of additional velocity
structure and supporting an AGN origin for much of the UV continuum. Overall,
the spectroscopic and continuum evidence converge on a picture in which a
compact AGN is embedded in a dense, partially covering gaseous cocoon.

\end{itemize}

Taken together, \target appears to be a textbook LRD: compact morphology,
exponential-like broad Balmer wings, narrow forbidden lines, multi-component
Balmer absorption, strong \FeII, and extreme gas densities, all pointing to an
AGN encased in stratified, high-pressure gas at early cosmic times. The results
have two immediate implications. First, caution is warranted when using broad-line
widths for \mbh in LRDs: plausible models alone introduce $\lesssim 0.6$~dex
systematics. Second, the combination of narrow forbidden-line kinematics and
strong BLR signatures argues for genuinely over-massive black holes rather than
for extraordinarily massive host galaxies inferred from stellar templates.

Future observations may use spatially resolved spectroscopy with higher spectral
resolution to map $\sigma$ versus critical density and position, firmly establishing
the stratified nature of the line-emitting regions. Higher-resolution spectroscopy
will also improve the identification and deblending of adjacent \FeII lines,
will provide more accurate \mdyn, and will independently assess \mbh via
dynamical probes \citep{juodzbalis+2025b}.
Deep, medium-resolution UV spectroscopy will confirm \FeIIperm absorption and
constrain its kinematics, enabling the study of feedback and testing
scattering-based AGN models.

Deeper rest-optical coverage to capture trans-auroral \OIIAuall and \OIL[8446]
at high SNR will be needed to assess the Fe/O ratio.

\begin{acknowledgments}

We acknowledge M. Begelman, V. Rusakov, M. Pettini, H. Katz, J. Scholtz and B. D. Johnson for useful discussions.
We further acknowledge V. Rusakov and G. Panagiotis Nikopoulos for sharing their unpublished analysis
of this source.
EJN gratefully acknowledges support from NASA through grant JWST-GO-04106,  awarded by the Space Telescope Science Institute, which is operated by the Association of Universities for Research in Astronomy, Inc., under NASA contract NAS 5-26555.
FDE, XJ and RM acknowledge support by the Science and Technology Facilities Council (STFC), by the ERC through Advanced Grant 695671 ``QUENCH'', and by the UKRI Frontier Research grant RISEandFALL.
RM also acknowledges funding from a research professorship from the Royal Society.

This work is based on observations made with the NASA/ESA/CSA James Webb Space
Telescope. The data were obtained from the Mikulski Archive for Space Telescopes
at the Space Telescope Science Institute, which is operated by the Association of
Universities for Research in Astronomy, Inc., under NASA contract NAS 5-03127 for
JWST. These observations are associated with program \#4106 and
\href{https://doi.org/10.17909/z7p0-8481}{\#1345 (CEERS)}.
The reduced JWST/NIRSpec data for \target are available from the authors upon reasonable request.
This work is based on observations made with the GTC telescope, in the Spanish Observatorio del Roque de los Muchachos of the Instituto de Astrof\'isica de Canarias, under Director's Discretionary Time.
The GTC spectrum of Lord of LRDs \citep{ji+2025b} is publicly available from zenodo \doi{10.5281/zenodo.17235199}.

The authors acknowledge the CEERS team for developing their observing program
with a zero-exclusive-access period.

\end{acknowledgments}

\begin{contribution}

EJN and FDE are co-first authors and performed the emission line analysis. XJ performed the \cloudy
modeling. JB performed the morphology and SED analysis.
All co-authors contributed to the interpretation of the analysis.

\end{contribution}

%
\facilities{JWST {NIRCam and NIRSpec}}

\software{
    Python \citep{vanrossum1995},
    \astropy \citep{2013A&A...558A..33A,2018AJ....156..123A,2022ApJ...935..167A},
    \cloudy \citep{2013RMxAA..49..137F},
    {\tt \href{https://pypi.org/project/corner/}{corner}} \citep{foreman-mackey2016},
    {\tt \href{https://sites.google.com/cfa.harvard.edu/saoimageds9}{ds9}} \citep{joye+mandel2003}.
    {\tt \href{https://pypi.org/project/emcee/}{emcee}} \citep{foreman-mackey+2013},
    {\tt \href{https://github.com/ryanhausen/fitsmap}{fitsmap}} \citep{hausen+robertson2022}
    {\tt \href{https://pypi.org/project/jwst/}{jwst}} \citep{alvesdeoliveira+2018},
    {\tt \href{https://pypi.org/project/matplotlib/}{matplotlib}} \citep{hunter2007},
    {\tt \href{https://pypi.org/project/numpy/}{numpy}} \citep{harris+2020},
    {\tt \href{https://pypi.org/project/PyNeb/}{pyneb}} \citep{luridiana+2015},
    {\tt \href{https://pypi.org/project/scipy/}{scipy}} \citep{jones+2001}.
}



\bibliography{irony}{}
\bibliographystyle{config/aasjournalv7}



\end{document}